\documentclass[fleqn,usenatbib]{mnras}

\usepackage{newtxtext,newtxmath}

\usepackage[T1]{fontenc}

\DeclareRobustCommand{\VAN}[3]{#2}
\let\VANthebibliography\thebibliography
\def\thebibliography{\DeclareRobustCommand{\VAN}[3]{##3}\VANthebibliography}

\usepackage{graphicx}	
\usepackage{amsmath}	
\usepackage{subfig}    
\usepackage{pgffor}  

\newcommand{\Lya}{{\rm Ly}$\alpha$ }
\defcitealias{mackenzie2021revealing}{M21}
\defcitealias{borisova2016ubiquitous}{B16}
\defcitealias{farina2019requiem}{F19}

\title[Ly$\alpha$ haloes of high-$z$ faint quasars]{Subaru High-$z$ Exploration of Low-Luminosity Quasars (SHELLQs). XXIII. The powering mechanisms of the Ly$\alpha$ haloes around high-$z$ quasars probed by slit spectroscopy}

\author[Hoshi et al.]{
Hiroki Hoshi,$^{1}$\thanks{E-mail: hirostar1013@gmail.com}
Rikako Ishimoto,$^{1}$
Nobunari Kashikawa,$^{1,2}$
Yoshiki Matsuoka,$^{3}$
Wanqiu He,$^{4}$
\newauthor
Junya Arita,$^{1}$
Kazushi Iwasawa,$^{5,6}$
Toshihiro Kawaguchi,$^{7}$
Satoshi Kikuta,$^{1}$
Rieko Momose,$^{1,8}$
\newauthor
Rhythm Shimakawa,$^{9}$
Shunta Shimizu,$^{1}$
Ayumi Takahashi,$^{4}$
Yoshihiro Takeda,$^{1}$
Yoshiki Toba,$^{3,4,10}$
\newauthor
Takehiro Yoshioka,$^{1}$
Chien-Hsiu Lee,$^{11}$and 
Yuri Nishimura$^{1,12}$
\\
$^{1}$Department of Astronomy, School of Science, The University of Tokyo, 7-3-1 Hongo, Bunkyo-ku, Tokyo 113-0033, Japan\\
$^{2}$Research Center for the Early Universe, The University of Tokyo, 7-3-1 Hongo, Bunkyo-ku, Tokyo, 113-0033, Japan\\
$^{3}$Research Center for Space and Cosmic Evolution, Ehime University, 2-5 Bunkyo-cho, Matsuyama, Ehime 790-8577, Japan\\
$^{4}$National Astronomical Observatory of Japan, 2-21-1 Osawa, Mitaka, Tokyo 181-8588, Japan\\
$^{5}$Institut de Ci\`encies del Cosmos (ICCUB), Universitat de Barcelona (IEEC-UB), Mart\'i i Franqu\`es, 1, 08028 Barcelona, Spain\\
$^{6}$ICREA, Pg Llu\'is Companys 23, 08010 Barcelona, Spain\\
$^{7}$Graduate School of Science and Engineering,
University of Toyama, Gofuku 3190, Toyama 930-8555, Japan\\
$^{8}$Observatories, Carnegie Science, 813 Santa Barbara Street, Pasadena,
CA 91101, USA\\
$^{9}$Waseda Institute for Advanced Study (WIAS), Waseda University, 1-21-1, Nishi-Waseda, Shinjuku, Tokyo 169-0051, Japan\\
$^{10}$Academia Sinica Institute of Astronomy and Astrophysics, 11F of Astronomy-Mathematics Building, AS/NTU, No.1, Section 4, Roosevelt Road, \\ Taipei 10617, Taiwan\\
$^{11}$Hobby-Eberly Telescope, McDonald Observatory, UT Austin
28 Mt Fowlkes Rd, Fort Davis, TX 79734, USA\\
$^{12}$Institute of Astronomy, The University of Tokyo,
2-21-1, Osawa, Mitaka, Tokyo 181-0015, Japan
}

\date{Accepted XXX. Received YYY; in original form ZZZ}

\pubyear{2024}

\begin{document}
\label{firstpage}
\pagerange{\pageref{firstpage}--\pageref{lastpage}}
\maketitle

\begin{abstract}
We present the analysis of Ly$\alpha$ haloes around faint quasars at $z\sim4$ and $z\sim6$. 
We use 20 and 162 quasars at $z\sim4$ and $z\sim6$, taken by slit spectroscopy, and detect Ly$\alpha$ haloes around 12 and 26 of these quasars, respectively. 
The average absolute magnitudes of the detected quasars are $\langle M_{1450} \rangle = -23.84$ mag at $z\sim4$ and $\langle M_{1450} \rangle = -23.68$ mag at $z\sim6$, which are comparable at $z\sim4$ and 3 mag fainter at $z\sim6$ than those of previous studies.
The median surface brightness profiles are found to be consistent with an exponential curve, showing a hint of flattening within a radius of 5 kpc.
The Ly$\alpha$ haloes around these faint quasars are systematically fainter than those around bright quasars in the previous studies.
We confirm the previous results that the Ly$\alpha$ halo luminosity depends on both the ionizing and Ly$\alpha$ peak luminosities of quasars at $z\sim4$, and also find that a similar correlation holds even at $z\sim6$. 
While the observed Ly$\alpha$ halo luminosity is overall attributed to recombination emission from the optically thin gas clouds in the CGM, its luminosity dependences can be consistently explained by the partial contributions of recombination radiation from the optically thick clouds, which is thought to originate from the CGM centre, and the scattered Ly$\alpha$ photons, which is resonantly trapped at the CGM centre and escaping outside of the haloes.
\end{abstract}




\section{Introduction}
More than 300 quasars have been detected at $z>6$~\citep{fan2023}, suggesting that supermassive black holes (SMBHs) with masses
$> 10^9\ \mathrm{M_{\sun}}$ have already formed by a cosmic age of 900 Myr. These observations pose a challenge to theoretical models, which should explain such rapid BH growth at early cosmic time. A massive $(\sim 10^5\ \mathrm{M_{\sun}})$ seed BH grows at a rate close to the Eddington limit, otherwise a much lower-mass $(\sim 10^2\ \mathrm{M_{\sun}})$ seed BH must sustain super-Eddington
accretion \citep[e.g.][]{volonteri2006}. In either case, the host galaxy of the $z\sim6$ quasar must retain a large amount of gas around it and not only consume it for its own active star formation but also supply it efficiently to the central BH. In the current $\Lambda$CDM paradigm of galaxy formation, high-$z$ quasars would reside in massive dark matter haloes ($\ga 10^{12-13}\mathrm{\ M_{\sun}}$
;~\citealp[e.g.][]{springel2005,costa2014}), which is supported by recent direct halo mass measurements \citep{arita2023,eilers2024}, in order to have sufficient gas to foster both BHs and stars~\citep{efsta1988}. Both of them consume large amounts of gas and must be constantly refilled with pristine gas by the inflow from the circum-galactic medium (CGM) and the intergalactic medium (IGM) \citep[e.g.][]{dekel2006}. Cosmological hydrodynamic simulation shows that steady high-density cold gas accretion is responsible for sustaining critical accretion rates leading to rapid growth of $\sim 10^9\mathrm{\ M_{\sun}}$ SMBHs as early as $z\sim7$ \citep[e.g.][]{matteo2012}. When the gas in the CGM around a quasar host galaxy is illuminated by ionizing radiation from the SMBH or by an intense starburst, it is observed as an extended ``\Lya halo'' (also referred to as \Lya nebula). 


The \Lya haloes around quasars are brighter and more extended than those of normal galaxies due to their strong AGN ionizing radiation in addition to the abundant CGM gas in the massive hosts~\citep{battaia2018}. In recent years, many surveys on \Lya haloes around high-$z$ quasars have been conducted using Integral-Field-Units (IFU) mounted on 8-m class telescopes, such as the VLT/Multi-Unit Spectroscopic Explorer (MUSE;~\citealt{bacon2010}) and the Keck Cosmic Web Imager (KCWI;~\citealt{morissey2018}).
\citet{borisova2016ubiquitous} (hereafter \citetalias{borisova2016ubiquitous}) revealed \Lya haloes from all 17 quasars at $z\sim3.5$, 
demonstrating that \Lya haloes, extending out to scales larger than 100 kpc, with clear asymmetries, are ubiquitously surrounding luminous quasars. 
\citet{2019MNRAS.482.3162A}  extended this to a survey of 61 quasars at $z\sim3$, showing that the \Lya haloes extend over tens of kpc and are characterised by relatively quiescent kinematics.
Of the few \Lya halo searches at $z\sim6$ \citep[e.g.][]{willott2011lyman,farina2017,momose2019,drake2019}, the most comprehensive survey is the Reionization
Epoch QUasar InvEstigation with MUSE (REQUIEM) survey (\citealt{farina2019requiem}; \citetalias{farina2019requiem}), detecting \Lya haloes around 12 from 31 quasar samples.
The REQUIEM survey shows that extended \Lya haloes are common around quasars even at $z\sim6$, with the characteristics consistent with $z\sim3$, suggesting that the same physical mechanism that produces the \Lya halo could be at work in both $z\sim3$ and $z\sim6$.
The \Lya halo provides enough gas to sustain rapid star formation and black hole growth in the early universe.

Although the properties of \Lya haloes are gradually being uncovered through these surveys, the mechanisms powering \Lya haloes remain unclear. The most commonly believed mechanism is the recombination radiation (also called \Lya fluorescence) from gas ionized by the quasar~\citep[e.g.][]{cantalupo2005,cantalupo2014,henwai2013,masribas2016}. In addition, the collisional excitation~\citep[e.g.][]{fardal2001,cantalupo2008,goerdt2010} and the resonant scattering of \Lya photons from the quasar~\citep[e.g.][]{cantalupo2017,dijkstra2017,gronke2017} are also suggested to contribute to the emission.
Different plausible mechanisms depend on different physical quantities for their emission, and their dependence on quasar luminosity varies accordingly. 
The \Lya halo luminosity is expected to be proportional to the quasar luminosity at the Lyman limit when recombination radiation from optically thick gas is effective, and to the quasar \Lya luminosity when resonant scattering is the main contributor~\citep[e.g.][]{henwai2013,hennawi2015,pezzuli2019}.
It is therefore crucial to expand the observed range of quasar luminosity to reliably constrain the emission mechanisms.
\citet{mackenzie2021revealing} (hereafter \citetalias{mackenzie2021revealing}) observed faint ($M_i \lesssim -23$) quasars at $z\sim3$ and showed a strong dependence of \Lya halo luminosity on quasar luminosity by comparing with \citetalias{borisova2016ubiquitous}. 
However, detection of faint quasars at higher redshift is still challenging because their surface brightness (SB) decreases rapidly in proportion to $(1+z)^{-4}$.
Neither quasar luminosity, star formation rate, nor SMBH mass dependence of \Lya halo luminosity is found at $z\sim6$ (\citealt{farina2019requiem,farina2022}), though this may be because the sample is biased toward luminous ones, and limited to very narrow luminosity ranges.
There have been no studies focusing on faint quasars comparable to \citetalias{mackenzie2021revealing} at $z\sim6$. 

In \Lya halo studies,
IFUs are often used because of their great advantage in providing spatially rich information. However, the observational cost of IFU is high, which makes it difficult to increase the number of samples. 
An alternative method is slit spectroscopy \citep{willott2011lyman,roche2014}, which is the most popular spectroscopy mode, and it benefits from the availability of numerous legacy datasets, allowing us to significantly increase the number of samples, though it provides only one-dimensional spatial information.
It also has the weakness of not being able to cover a large area; therefore, it has a poor detection ability especially outside of a large halo.
On the other hand, the IFU has the difficulty that the point spread function (PSF) in the reconstructed cube is complex and cannot be easily modelled in the analysis profile \citepalias{borisova2016ubiquitous}, making it difficult to explore inside the halo.
The superiority of IFU over slit remains the same when observing the \Lya haloes, which generally show asymmetric morphology; 
however, slit spectroscopic data is still useful.
Recent James Webb Space Telescope (JWST) observations with Near Infrared Spectrograph (NIRSpec) MSA have also detected extended \Lya emission even at $z>10$, suggesting the presence of \Lya haloes at such high redshifts~\citep{bunker2023,maiolino2024,scholtz2024}.

In this study, we attempt to search for \Lya halo by collecting slit spectroscopy datasets, especially for faint quasars. 
We aim to place constraints on the mechanisms powering \Lya haloes by combining these results with those from previous studies of brighter samples, \citetalias{borisova2016ubiquitous} at $z\sim4$ and \citetalias{farina2019requiem} at $z\sim6$. 
The increase in \Lya halo data around faint quasars obtained through slit spectroscopy will contribute to the progress in \Lya halo research.

This paper is structured as follows. Section~\ref{sec:sample} describes the datasets and the process leading to the detection of \Lya haloes. The resulting SB radial profiles and luminosity dependence are discussed in Section~\ref{sec:results}, and we discuss the powering mechanisms of \Lya haloes in Section~\ref{sec:discussion}.
Throughout this paper, we use the AB magnitude system~\citep{oke1983}. We assume a standard $\Lambda$CDM cosmology with $h = 0.7,\ \Omega_m = 0.3,\ \Omega_\Lambda = 0.7$.
We use pkpc to indicate the physical scale.

\section{THE SAMPLE}\label{sec:sample}
\subsection{The quasar sample at $z \sim 4$}
The $z\sim4$ spectroscopic sample consists of 20 quasars at $3.45 \leq z \leq 4.10$ taken by \citet{he2024black}. 
These are low-luminosity ($20 < m_i < 24$, where $m_i$ is the $i$ band magnitude) $z \sim 4$ quasars selected by using the data set~\citep{akiyama2018quasar} of the Hyper Suprime-Cam Subaru Strategic Program (HSC-SSP; \citealt{aihara2018hyper,aihara2018first,aihara2019second,aihara2022third}) .
HSC-SSP is a wide-field multi-band imaging survey utilising HSC~\citep{miyazaki2018hyper}, a wide-field CCD camera attached at the prime focus of the Subaru 8.2 m telescope. 
Based on the S16A-Wide2 data release, \citet{akiyama2018quasar} construct a sample of 1666 $z \sim 4$ low-luminosity quasar candidates in an effective area
of $172\ \mathrm{deg^2}$ mainly with colour selections, and \citet{he2024black} imposed additional colour criteria to recover quasar candidates that are missed in \citet{akiyama2018quasar}. 
\citet{he2024black} conducted follow-up spectroscopic observations for 2361 candidates using the two-degree field (2dF) fibre positioner (\citealt{lewis2002}) with the AAOmega spectrograph (\citealt{smith2004}) mounted on the 3.9m Anglo-Australian Telescope (AAT) and the DEep Imaging Multi-Object
Spectrograph (DEIMOS;~\citealt{faber2003deimos}) mounted on the Keck II telescope. We only use 20 samples that have been spectroscopically identified by DEIMOS, because the fibre-fed spectrograph AAT/AAOmega is not suitable for our analysis. 
The DEIMOS data is taken with the 1$\arcsec$ slit, 600ZD grating, and the GG495 blue-blocking filter.
Seeing is mostly $0.5\text{--}0.7\arcsec$. 
The pixel scale is 0.1185$\arcsec$ pixel$^{-1}$ and the typical wavelength coverage is $\sim 5000\text{--}10000$ \AA $\ $with a resolution of $R\sim1600$. 
Each object is observed with a total exposure time of 40--90 minutes, typically divided into 2--3 exposures of 20--30 minutes. See \citet{he2024black} for detailed sample selections and observations.
We should note that these data are acquired to measure BH mass, and the integration time is not optimized for detecting Ly$\alpha$ haloes.
We use reduced two-dimensional spectral data with a spatial extent of 9.4$\arcsec$.
The summary of $z\sim4$ quasar sample is shown in Table~\ref{tab:z4sample}.

\begin{table*}
	\centering
	\caption{Properties of our $z\sim4$ sample}
	\label{tab:z4sample}
	\begin{tabular}{lcccccc} 
		\hline
		ID & R.A. & Decl. & Redshift & $M_{1450}$ (mag) & Exp. Time (min) & detected$^{(1)}$\\
		\hline
J0914$-$0125 & 09:14:08.19 & $-$01:25:11.9 & 3.94 & $-23.60$ & 90 & d\\
J0925$+$0239 & 09:25:53.15 & $+$02:39:56.3 & 3.53 & $-24.00$ & 60 & d\\
J1156$-$0108 & 11:56:08.96 & $-$01:08:53.4 & 3.36 & $-25.10$ & 60 & d\\
J1156$-$0053 & 11:56:12.31 & $-$00:53:18.9 & 4.10 & $-26.07$ & 60 & \\
J1156$-$0059 & 11:56:16.98 & $-$00:59:37.0 & 3.76 & $-22.87$ & 60 & d\\
J1156$-$0058 & 11:56:26.33 & $-$00:58:32.9 & 3.78 & $-23.22$ & 60 & d\\
J1203$+$0032 & 12:03:57.94 & $+$00:32:05.1 & 3.73 & $-24.64$ & 60 & d\\
J1204$+$0032 & 12:04:34.27 & $+$00:32:24.7 & 3.97 & $-23.95$ & 60 & \\
J1415$+$0057 & 14:15:41.74 & $+$00:57:20.6 & 3.94 & $-24.19$ & 60 & \\
J1415$+$0103 & 14:15:50.76 & $+$01:03:00.2 & 3.79 & $-23.40$ & 60 & d\\
J1418$+$0106a & 14:18:12.46 & $+$01:06:04.4 & 3.70 & $-23.16$ & 60 & d\\
J1418$+$0106b & 14:18:38.72 & $+$01:06:01.9 & 3.54 & $-24.62$ & 60 & d\\
J1418$+$0107 & 14:18:44.59 & $+$01:07:47.0 & 3.67 & $-23.28$ & 60 & d\\
J1447$-$0131 & 14:47:42.68 & $-$01:31:40.4 & 3.79 & $-23.14$ & 70 & \\
J1448$-$0142 & 14:48:03.28 & $-$01:42:19.3 & 3.68 & $-23.59$ & 70 & d\\
J1448$-$0145 & 14:48:05.19 & $-$01:45:47.5 & 3.63 & $-23.34$ & 70 & \\
J1557$+$4211 & 15:57:00.21 & $+$42:11:53.4 & 3.94 & $-22.34$ & 40 & \\
J1613$+$4203 & 16:13:53.55 & $+$42:03:30.7 & 4.04 & $-23.95$ & 40 & \\
J1634$+$4300 & 16:34:13.28 & $+$43:00:31.3 & 3.92 & $-24.62$ & 40 & d\\
J1634$+$4251 & 16:34:42.97 & $+$42:51:04.4 & 3.45 & $-22.91$ & 40 & \\
		\hline
	\end{tabular}\\
 \emph{NOTE} (1) The letter ``d'' is indicated if the \Lya halo is detected.
\end{table*}

As will be discussed later, \Lya haloes are detected in 12 
quasars.
The detected sample has an average redshift of $\langle z \rangle = 3.70$ and an average absolute magnitude at 1450 \AA $\ $of $\langle M_{1450} \rangle = -23.84$ mag. 
Following \citetalias{mackenzie2021revealing}, the conversion from $m_i$ to $M_i$ is given by the following equation,
\begin{equation}
    M_i = m_i-5\log{100d_\mathrm{L}}+2.5(1+\alpha_\nu)\log{(1+z)} \label{eq:mi2Mi}
\end{equation}
where $d_\mathrm{L}$ is the luminosity distance in the unit of kpc and $\alpha_\nu$ is the
power-law index of the continuum ($f_\nu \propto \nu^{\alpha_\nu}$). 
We adopt $\alpha_\nu=-0.5$ \citep{2006AJ....131.2766R} in our calculations following \citetalias{mackenzie2021revealing}.
Then, the $M_i$ is converted to $M_{1450}$, using
\begin{equation}
    M_{1450} = M_i+0.684
    \label{eq:Mi2M1450}.\\
\end{equation}
This relationship is empirically determined by \citet{2015MNRAS.449.4204L} based on a sample of $z\sim2.4$ quasars, but we assume that this also holds at $z\sim4$ because the SED of quasars does not evolve much \citep[e.g.][]{fan2023}.

\begin{figure}
	\includegraphics[width=\columnwidth]{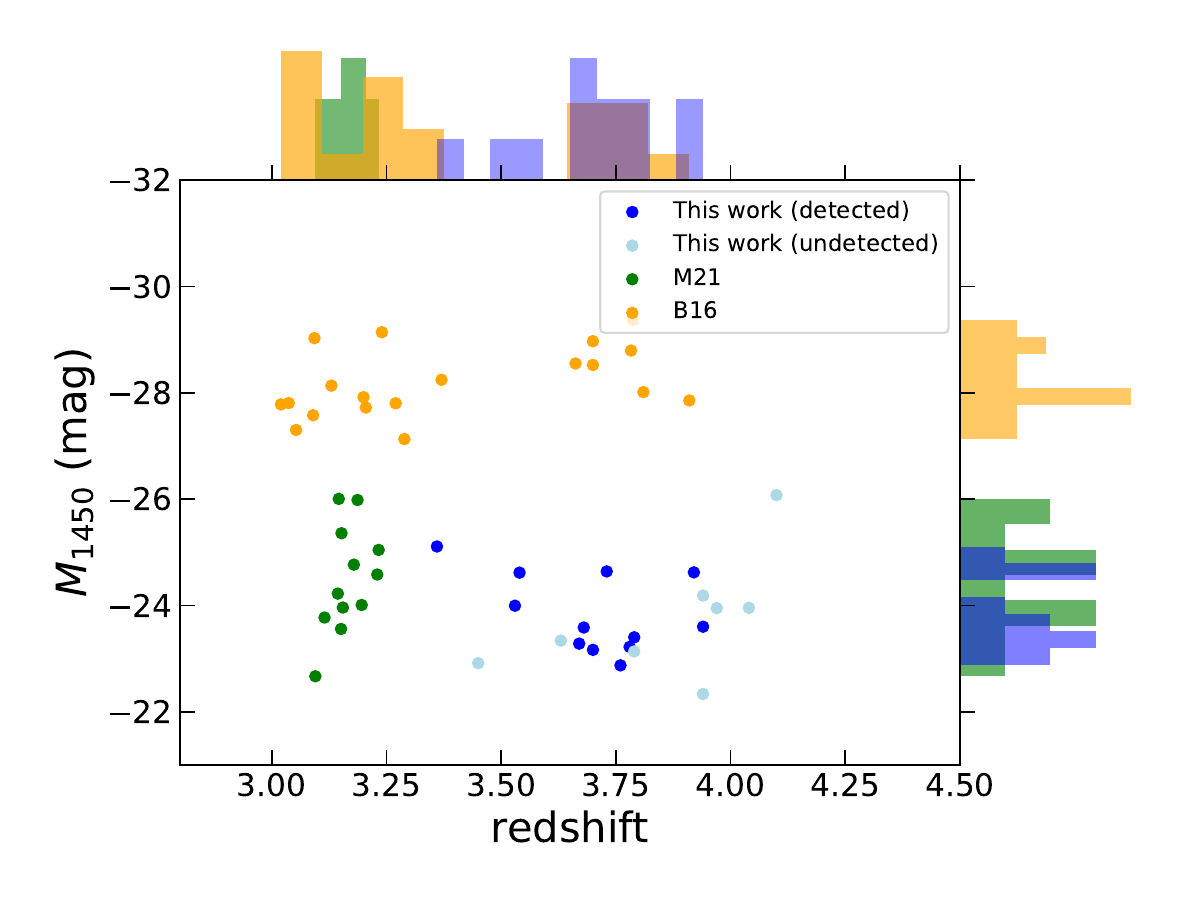}
    \caption{The redshift and magnitude distribution of \citetalias{borisova2016ubiquitous} (orange), \citetalias{mackenzie2021revealing} (green), and our $z\sim4$ (blue) quasars. Quasars with no \Lya halo detection are also shown in light blue. \Lya haloes are detected around all quasars for \citetalias{borisova2016ubiquitous} and \citetalias{mackenzie2021revealing}. The histograms at the top and right of the figure represent the redshift and $i$-band absolute magnitude distributions of the sample with the \Lya halo detection, respectively. The histograms are normalised so that the total is one.}
    \label{fig:z4zM}
\end{figure}

Figure~\ref{fig:z4zM} shows the redshift and $M_{1450}$ distributions of the $z\sim4$ sample, compared with the previous studies that detected the \Lya halo with MUSE, \citetalias{borisova2016ubiquitous}, which observed 19 bright quasars at $z \sim 3.5$,  and \citetalias{mackenzie2021revealing}, which observed 12 faint quasars at $z \sim 3.15$. 
Our sample is fainter than \citetalias{borisova2016ubiquitous} and comparable to \citetalias{mackenzie2021revealing}.
Our sample has a slightly higher redshift than the \citetalias{mackenzie2021revealing} sample, but it can be assumed that the effect of the redshift difference between the two is small.


\subsection{The quasar sample at $z \sim 6$}
Our $z \sim 6$ sample comes from Subaru High-z Exploration of Low-Luminosity Quasars project (SHELLQs), which consists of 162 spectroscopically confirmed quasars at $5.66 \leq z \leq 7.07$ \citep{matsuoka2016subaru,matsuoka2018subaru,matsuoka2018subaru2,matsuoka2019discovery,matsuoka2019subaru,matsuoka2022subaru}. 
The quasar candidates are selected from the HSC-SSP data by colour criteria and the Bayesian algorithm, and spectroscopic identification is carried out with the Subaru/Faint Object Camera and Spectrograph (FOCAS; \citealt{kashikawa2002focas}) or the GTC/Optical System for Imaging and low-intermediate-Resolution Integrated Spectroscopy (OSIRIS; \citealt{cepa2000osiris}).  See \citet{matsuoka2016subaru,matsuoka2018subaru,matsuoka2018subaru2,matsuoka2019discovery,matsuoka2019subaru,matsuoka2022subaru} for detailed selection. 
The FOCAS data is taken by the MOS mode with the VPH900 grism and SO58 order-sorting filter. 
The slit width is 1$\arcsec$, and the seeing size is typically 0.4--1.2$\arcsec$. 
The spatial pixel scale is 0.208$\arcsec$ pixel$^{-1}$ and the wavelength coverage is 0.75--1.05 $\mu$m with a resolution of $R\sim1200$.
The OSIRIS data is taken with the R2500I grism. 
The slit width is 1$\arcsec$, and the seeing size is typically 0.7--1.3$\arcsec$. 
The pixel scale is 0.254$\arcsec$ pixel$^{-1}$ and the wavelength coverage is 0.74--1.0 $\mu$m with a resolution of $R\sim1500$.
Note that these spectroscopic data are acquired primarily for quasar identification, and many of them are taken with short exposures, and like $z\sim4$ samples, they have not been integrated long enough to detect Ly$\alpha$ halo.
We use reduced two-dimensional spectral data with a spatial extent of 6.8$\arcsec$.
\begin{figure}	\includegraphics[width=\columnwidth]{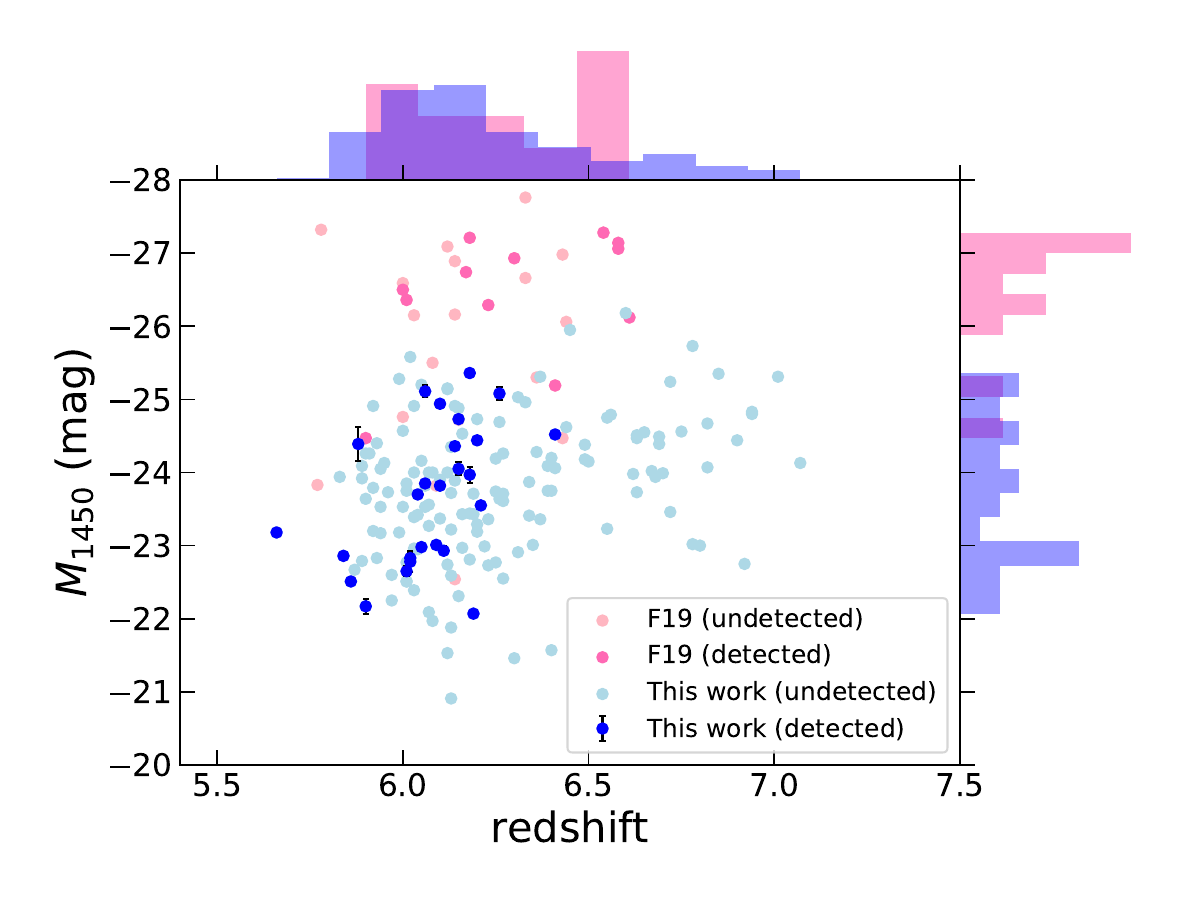}
    \caption{The redshift and magnitude distributions of $z\sim6$ quasars. Our sample is shown in blue, while \citetalias{farina2019requiem} is shown in pink. For both samples, quasars with detected \Lya haloes are represented in darker colours, and those without detections are shown in lighter colours. The histograms at the top and right of the figure represent the redshift and $M_{1450}$ distributions of the sample with the \Lya halo detection, respectively. The histograms are normalised so that the total is one.}
    \label{fig:z6zM}
\end{figure}

Figure~\ref{fig:z6zM} shows the redshift and $M_{1450}$ distributions of the $z\sim6$ \Lya halo-detected sample (see Section \ref{sec:detection}) and not-detected sample, compared with \citetalias{farina2019requiem}, a representative \Lya halo survey at $z \sim 6$. $M_{1450}$ is given in suites of SHELLQs papers and \citetalias{farina2019requiem}.
The halo-detected sample has an average redshift of $\langle z \rangle = 6.07$ and an average absolute magnitude of $\langle M_{1450} \rangle = -23.68$ mag.
This sample is notably fainter than \citetalias{farina2019requiem} and makes it possible to extend to the low-luminosity regime in \Lya halo studies at this redshift. 

\begin{table*}
	\centering
	\caption{Properties of our Ly$\alpha$ halo-detected sample of $z\sim6$ quasars}
	\label{tab:z6sample}
\begin{tabular}{lccccccr} 
		\hline
		ID & R.A. & Decl. & Redshift & $M_{1450}$ (mag) & Exp. Time (min) & Instrument$^{(1)}$ & Paper$^{(2)}$\\
		\hline
J0122-0036 & 01:22:35.47 & $-$00:36:02.4 & 6.26 & $-25.08\pm0.09$ & 15 & O & XVI\\
J0220-0432 & 02:20:29.71 & $-$04:32:03.9 & 5.90 & $-22.17\pm0.10$ & 70 & F & IV\\
J0235-0532 & 02:35:42.42 & $-$05:32:41.6 & 6.09 & $-23.01\pm0.05$ & 60 & F & II\\
J0834+0211 & 08:34:00.88 & $+$02:11:46.9 & 6.15 & $-24.05\pm0.09$ & 40 & O & IV\\
J0844-0132 & 08:44:08.61 & $-$01:32:16.5 & 6.18 & $-23.97\pm0.11$ & 60 & O & IV\\
J0844+0423 & 08:44:22.57 & $+$04:23:53.7 & 6.21 & $-23.55\pm0.05$ & 15 & F & XVI\\
J0957+0053 & 09:57:40.40 & $+$00:53:33.7 & 6.05 & $-22.98\pm0.04$ & 75 & F & IV\\
J1004+0239 & 10:04:01.37 & $+$02:39:30.9 & 6.41 & $-24.52\pm0.03$ & 30 & F & IV\\
J1020+0429 & 10:20:47.40 & $+$04:29:46.7 & 6.18 & $-25.36\pm0.02$ & 15 & O & XVI\\
J1028+0017 & 10:28:41.66 & $+$00:17:55.9 & 6.10 & $-24.94\pm0.02$ & 15 & F & XVI\\
J1037+0037 & 10:37:34.52 & $+$00:37:50.8 & 6.11 & $-22.93\pm0.06$ & 60 & O & XVI\\
J1107-0118 & 11:07:56.01 & $-$01:18:19.0 & 6.06 & $-25.11\pm0.08$ & 15 & O & XVI\\
J1132+0038 & 11:32:18.15 & $+$00:38:00.1 & 5.66 & $-23.18\pm0.05$ & 100 & F & X\\
J1201+0133 & 12:01:03.02 & $+$01:33:56.4 & 6.06 & $-23.85\pm0.02$ & 120 & F & II\\
J1202+0256 & 12:02:53.13 & $+$02:56:30.8 & 6.02 & $-22.78\pm0.14$ & 15 & F & XVI\\
J1209-0006 & 12:09:24.01 & $-$00:06:46.5 & 5.86 & $-22.51\pm0.05$ & 60 & F & IV\\
J1317+0127 & 13:17:32.73 & $+$01:27:41.6 & 5.88 & $-24.39\pm0.23$ & 15 & O & XVI\\
J1347-0157 & 13:47:33.69 & $-$01:57:50.6 & 6.15 & $-24.73\pm0.02$ & 15 & F & X\\
J1400-0125 & 14:00:30.00 & $-$01:25:20.9 & 6.04 & $-23.70\pm0.05$ & 50 & F & IV\\
J1417+0117 & 14:17:28.67 & $+$01:17:12.4 & 6.02 & $-22.83\pm0.05$ & 60 & F & II\\
J1448+4333 & 14:48:23.33 & $+$43:33:05.9 & 6.14 & $-24.36\pm0.04$ & 30 & O & X\\
J1512+4422 & 15:12:48.71 & $+$44:22:17.5 & 6.19 & $-22.07\pm0.04$ & 30 & F & X\\
J1550+4318 & 15:50:00.93 & $+$43:18:02.8 & 5.84 & $-22.86\pm0.04$ & 45 & F & IV\\
J2216-0016 & 22:16:44.47 & $-$00:16:50.1 & 6.10 & $-23.82\pm0.04$ & 30 & F & I\\
J2228+0128 & 22:28:27.83 & $+$01:28:09.5 & 6.01 & $-22.65\pm0.07$ & 80 & F & I\\
J2255+0503 & 22:55:20.78 & $+$05:03:43.3 & 6.20 & $-24.44\pm0.02$ & 30 & O & X\\
		\hline
	\end{tabular}\\
 \emph{NOTE}
 (1): The letters ``F'' and ``O'' denote Subaru/FOCAS and GTC/OSIRIS, respectively. \\(2):
 I: \citet{matsuoka2016subaru}, II: \citet{matsuoka2018subaru},
 IV: \citet{matsuoka2018subaru2},
 X: \citet{matsuoka2019subaru},
 XVI: \citet{matsuoka2022subaru}
\end{table*}

\subsection{Quasar PSF subtraction}
Following \citet{willott2011lyman}, we carefully estimate the PSF of a quasar at each wavelength and subtract it from the observed two-dimensional spectrum to extract the \Lya halo. 
We estimate the PSF by assuming that it can be approximated as a Gaussian function in the spatial direction. 
The procedure is as follows.

\begin{figure}
\includegraphics[width=\columnwidth]{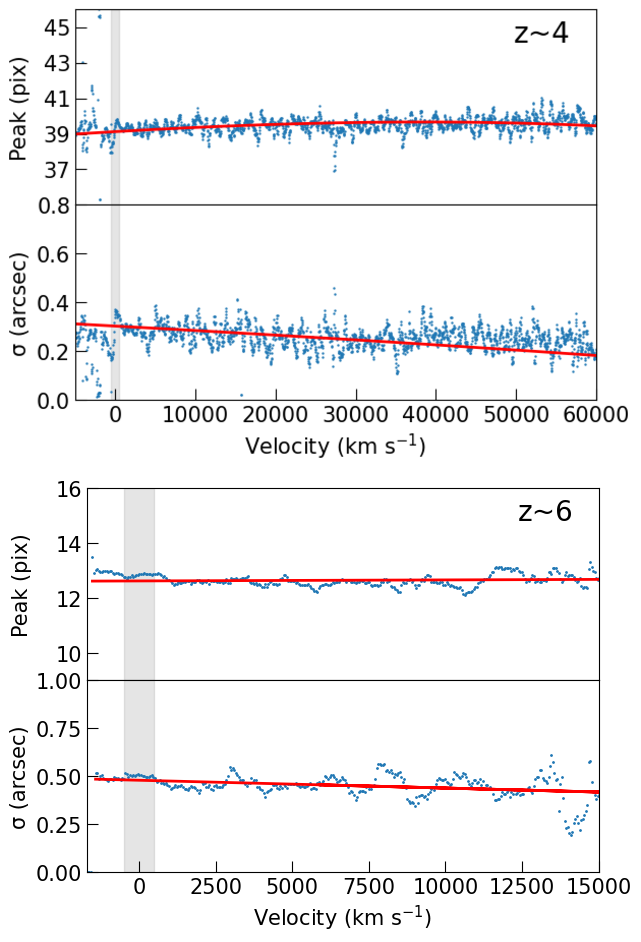}
    \caption{An example of the Gaussian fitting results for $z\sim4$ (top) and $z\sim6$ (bottom), respectively. The location of the centre of the Gaussian is shown in the upper panels and the Gaussian $\sigma$ is shown in the lower panels. The results of the linear or quadratic fits are also shown as red lines. The range from $-500$ to $+500\ \mathrm{km\ s^{-1}}$ (grey shaded) is not used for the fitting.}
    \label{fig:z46estimate}
\end{figure}

Firstly, we perform four ($\sim2.5\ $\AA) and 20 ($\sim30\ $\AA) pixel binning along the wavelength for $z\sim4$ and $z\sim6$, respectively, by taking the moving average to improve the signal-to-noise ratio (SNR). 
For $z\sim6$, we mask the wavelengths corresponding to strong sky emissions outside the range of $-500$ to $+500$ km s$^{-1}$ from the \Lya emission peak. 
Secondly, Gaussian fitting is performed for each wavelength bin to determine spatial peak location and Gaussian standard deviation, $\sigma$ of the unresolved AGN-dominated component. 
After outliers are excluded by 3$\sigma$-clipping, these parameters are estimated by fitting a linear or quadratic function using $\chi^2$ minimisation outside the \Lya halo wavelength regions, $-500$ to $+500\ \mathrm{km\ s^{-1}}$ from the \Lya emission peak (grey shaded region in Figure~\ref{fig:z46estimate}), of the spectrum. The centre position and $\sigma$ at the \Lya halo wavelengths 
are determined by extrapolating the fit results.
A linear fit is generally used, but for some $z\sim4$ samples with optically distorted two-dimensional spectra, a quadratic function fit is used for peak estimation instead of a linear fit.
Figure~\ref{fig:z46estimate} shows how the Gaussian centre and $\sigma$ are estimated.
At $z\sim 4$, where a large number of pixel data is available for a fitting and the wavelength range is wide, the continuum is clearly detected. 
While in the case of $z\sim 6$, the data available for fitting is limited to the red side of \Lya due to IGM absorption. 
In this way, the centre and extent of the PSF are estimated for each wavelength.

\begin{figure}	\includegraphics[width=\columnwidth,height=6cm]{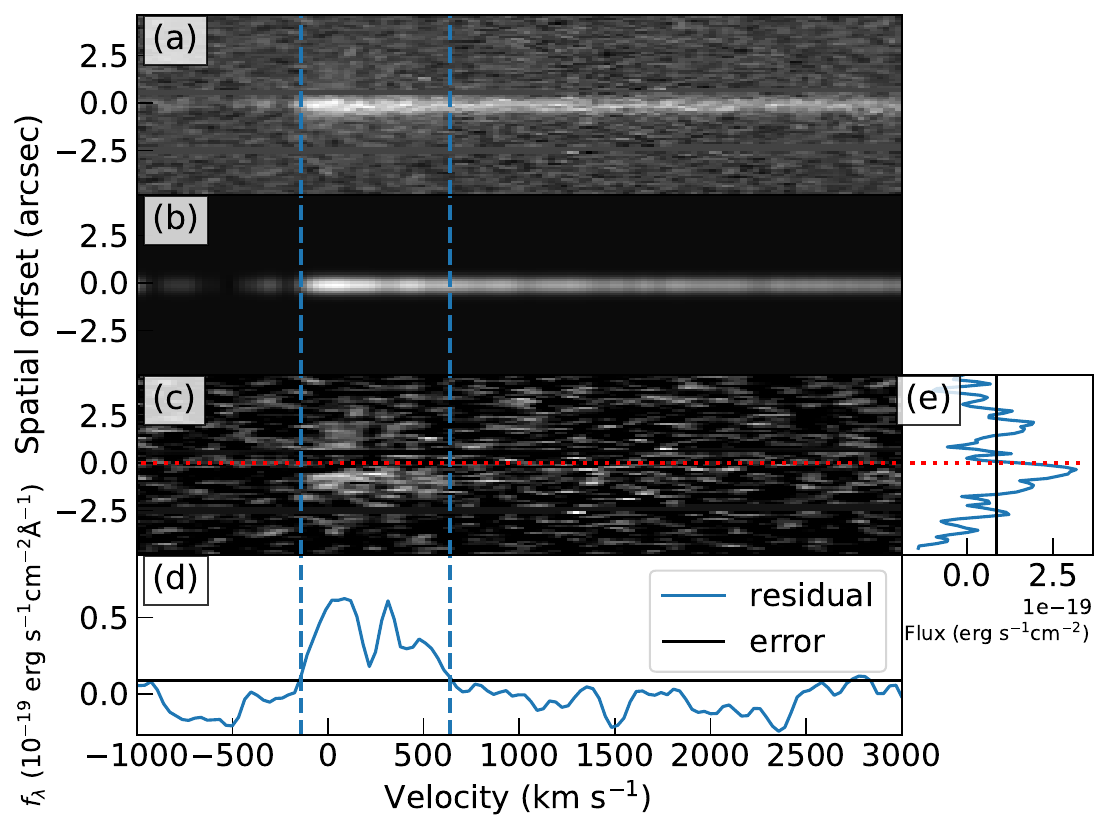}
    \caption{An example of the PSF subtraction at $z\sim 4$.
    (a) The original two-dimensional spectrum, (b) the PSF, (c) the residual \Lya emission, (d) the one-dimensional spectrum integrating over the spatial, and (e) wavelength directions. \Lya halo is detected as the residual, 
    which is shown as the blue solid line in (d) and (e). The region between two blue dashed lines corresponds to the wavelengths where the SNR calculated after the first PSF subtraction is larger than two to integrate the residual fluxes. The error is evaluated from the background variance (dotted line). The red dotted lines in (c) and (e) indicate the spatial centre, determined by averaging the Gaussian centres estimated as a function of wavelength, as shown in Figure~\ref{fig:z46estimate}, at corresponding wavelengths between the two blue dashed lines. The offset between the red dotted line and the peak of the one-dimensional spectrum could be due to the asymmetry of the \Lya halo.
    }
    \label{fig:z4sub}
\end{figure}

\begin{figure}	\includegraphics[width=\columnwidth,height=6cm]{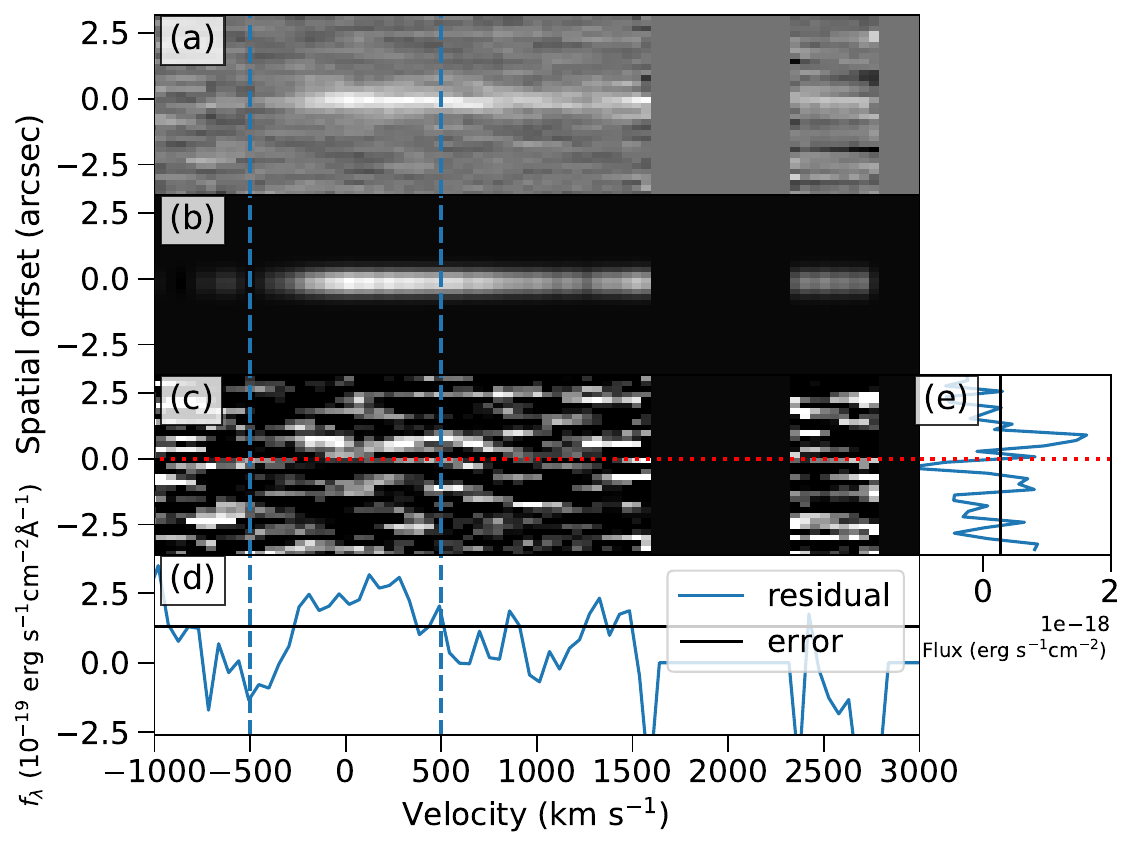}
    \caption{Same as Figure~\ref{fig:z4sub} but 
    at $z\sim 6$. The region between two blue dashed lines corresponds to the wavelength with $-500<v<500\ \mathrm{km\ s^{-1}}$. In panel (c), the contrast is adjusted so that the residual flux is easier to see. 
    }
    \label{fig:z6sub}
\end{figure}

Thirdly, the amplitude of the PSF to be subtracted is estimated. 
From here, two ($\sim 1.3$ \AA) and five ($\sim7.5$ \AA) pixel binning along the wavelength are used for $z\sim4$ and $z\sim6$, respectively. 
This is done to increase the SNR, but to avoid excessive degradation of spectral wavelength resolution, the binning width is set smaller than that used in the PSF parameter estimation described above.
The PSF amplitude is determined by performing Gaussian fitting again, fixing the centre position and $\sigma$ for each wavelength bin with the amplitude as the only free parameter. 
However, this method may overestimate the PSF amplitude around the \Lya wavelength, because, in principle, the PSF amplitude should be the sum of the quasar and the \Lya halo. 
Therefore, we perform a double-Gaussian fit for $z\sim 4$ objects, which have sufficient SNR, 
assuming that the extended \Lya halo component can also be approximated by a Gaussian. 
The double-Gaussian fits are performed at wavelengths where the SNR after the first PSF subtraction is larger than two.
For the SNR calculation, the signal is measured by summing the residuals after the quasar-PSF subtraction within the 2$\arcsec$, while the corresponding noise is evaluated from the background variance.
For $z\sim6$ objects, we use a single Gaussian fit, because double-Gaussian fit is not feasible due to the lack of SNR.
Finally, the halo components can be extracted by subtracting the quasar PSF from the original two-dimensional spectra. 
Figure~\ref{fig:z4sub},\ref{fig:z6sub} show examples of the results of PSF subtraction and \Lya halo detection for the sample at $z\sim4$ and $z\sim6$. 

The actual PSF might deviate from the Gaussian due to the PSF wings and a contribution from the host galaxy. 
To evaluate the validity of assuming the PSF to be a Gaussian, we quantitatively assess a possible under- or over-subtraction of the PSF.
On the PSF-subtracted image, we compare the median SB of the inner part (within $3\sigma$ of the Gaussian distribution), where the PSF is effectively subtracted, with that of the outer part, which is far away separated from the centre.
This comparison is made at the sufficiently long wavelengths of $v>2500\ \mathrm{km\ s^{-1}}$ far from the \Lya halo.
The median SB in the inner part is systematically smaller by $\sim10^{-16}\ \mathrm{erg\ s^{-1}cm^{-2}arcsec^{-2}}$ than that of the outer part, and the difference is about $\sim2$ dex smaller than the SB of the detected \Lya halo. 
We therefore conclude that the deviation of the PSF from the Gaussian is negligible.
The exponential law may be a better choice than Gaussian to approximate the spatial profile of \Lya halo (see Section~\ref{sec:radial}). 
We attempt to replace a Gaussian with an exponential model for the halo profile. We confirm that the results remain the same, though the inner diameter, around 3.3 pkpc, where halo SB dominates the PSF becomes slightly larger.
Furthermore, our method with a double-Gaussian might underestimate the quasar PSF amplitude and consequently overestimate the \Lya halo flux compared to the previous works of \citetalias{borisova2016ubiquitous},\citetalias{mackenzie2021revealing}, and \citetalias{farina2019requiem}, which determines the PSF amplitude at the centre of the original spectrum. 
To quantify this effect, we compare the results of the same analysis using a single Gaussian to the $z\sim4$ sample. 
When using a single Gaussian, the amplitude of PSF becomes higher than when using a double-Gaussian, so the inner diameter where halo SB dominates the PSF becomes larger, around 4.3 pkpc, and the SB profile (Section~\ref{sec:radial}) does not change outside of this.
The estimated \Lya halo luminosity (Section~\ref{sec:ldep}) is affected by about $\pm20\%$ at most, which is within its 1$\sigma$ error. 
Therefore, using a double-Gaussian allows us to measure the SB profile effectively even slightly inside the halo, but it does not significantly change the halo luminosity.



\subsection{Detection of Ly$\alpha$ halo}\label{sec:detection}
We measure the \Lya halo flux at $z\sim 4$ by integrating the residual fluxes within 2$\arcsec$ in the wavelength range where SNR of the one-dimensional spectrum after the quasar-PSF subtraction is larger than two, following the same threshold with \citetalias{mackenzie2021revealing} and \citetalias{borisova2016ubiquitous}.
On the other hand for $z\sim 6$, the \Lya halo flux is measured by just summing the residual fluxes between $-500$ and $+500\ \mathrm{km\ s^{-1}}$ from the \Lya emission peak, following \citetalias{farina2019requiem}.
Although we use slightly different wavelength ranges at $z\sim 4$ and $z\sim 6$, these are determined following previous studies to be carefully compared at each redshift.
The average integrated wavelength width at $z\sim4$ is about 28 \AA $\ $in the observed frame, which roughly corresponds to a wavelength width of about 30 \AA $\ $corresponding to the velocity range at $z\sim6$, but the central wavelength is not necessarily $0\ \mathrm{km\ s^{-1}}$.
The \Lya halo detection criterion is that its SNR, based on the residual fluxes within 2$\arcsec$ from the centre and the noise estimated from the background variance, exceeds the threshold of three. 
The SNR criteria for our integral \Lya flux might be lower than those of \citetalias{mackenzie2021revealing} and \citetalias{borisova2016ubiquitous}, which use voxels with SNR$>2$; however, it is challenging to perfectly match both criteria, which differ in instrument, methodology, and spatial extent of the halo under study.
All \Lya haloes are visually confirmed not to be fake.

\begin{figure}	\includegraphics[width=\columnwidth,height=6cm]{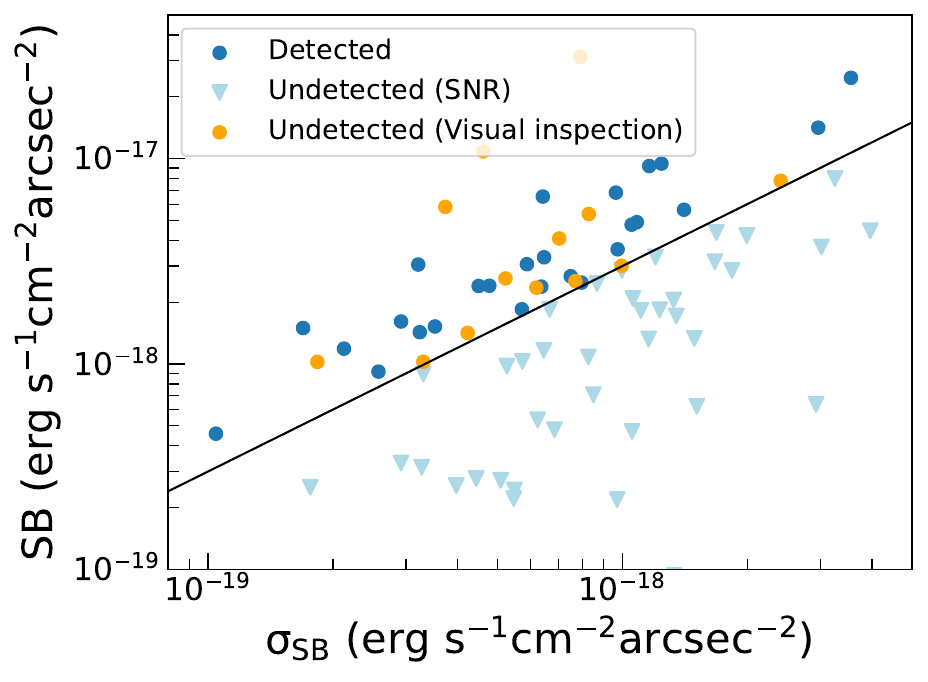}
    \caption{The average SB as a function of the SB variance  within a radius of 2$\arcsec$ in the $-500$ and $+500\ \mathrm{km\ s^{-1}}$ range for the $z\sim6$ sample. 
    Objects with \Lya halo detection are shown as blue circles, non-detections based on SNR are shown as light blue triangles, and those removed through visual inspection are shown as orange circles. The black line represents the 3$\sigma$ detection threshold.}
    \label{fig:z6snr}
\end{figure}

Finally, 12 and 26 haloes are detected at $z\sim 4$ and $z\sim 6$, respectively. 
The objects in which \Lya halo is detected at $z\sim4$ are marked with ``d'' in Table~\ref{tab:z4sample}, and the summary of the \Lya halo-detected sample of $z\sim6$ quasars is listed in Table~\ref{tab:z6sample}.
At $z\sim4$, \Lya haloes are not detected for eight objects, and these are all cases where the PSF fitting cannot be performed due to stray light, artificial effects, or the continuum spectrum being too-faint.
At $z\sim6$, we have 136 halo-non detected objects, and after excluding 50 objects for which PSF fitting is not possible for the same reason in the $z\sim4$ sample, there are 73 objects with SNR $<3$, and 13 objects are removed through visual inspection. 
Figure~\ref{fig:z6snr} shows the average SB as a function of the SB variance within a radius of 2$\arcsec$ in the $-500$ and $+500\ \mathrm{km\ s^{-1}}$ range for the $z\sim6$ sample. 
The typical SB depths used for SNR calculation are $2.9\times10^{-19}$ and $5.2\times10^{-19}\ \mathrm{erg\ s^{-1}cm^{-2}arcsec^{-2}}$ at $z\sim4$ and $z\sim6$, respectively. 
In addition, the one-dimensional spectra of PSF-subtracted images of halo-detected and halo-non detected objects are shown in Appendix~\ref{full_oned}.

It should be noted that in the analysis of long-slit spectroscopic data, where only one-dimensional spatial direction can be traced for a spatially asymmetric quasar \Lya halo, it is inevitable that the results of whether or not a \Lya halo is detected depend on the slit direction.
The fact that no \Lya halo is detected in a quasar in this study does not necessarily mean that there is no \Lya halo above a limiting SB in the quasar.
In this sense, the detection of \Lya halo in this analysis is not complete. 

\section{RESULTS} \label{sec:results}
\subsection{Radial surface brightness profiles}\label{sec:radial}
\begin{figure}	\includegraphics[width=\columnwidth]{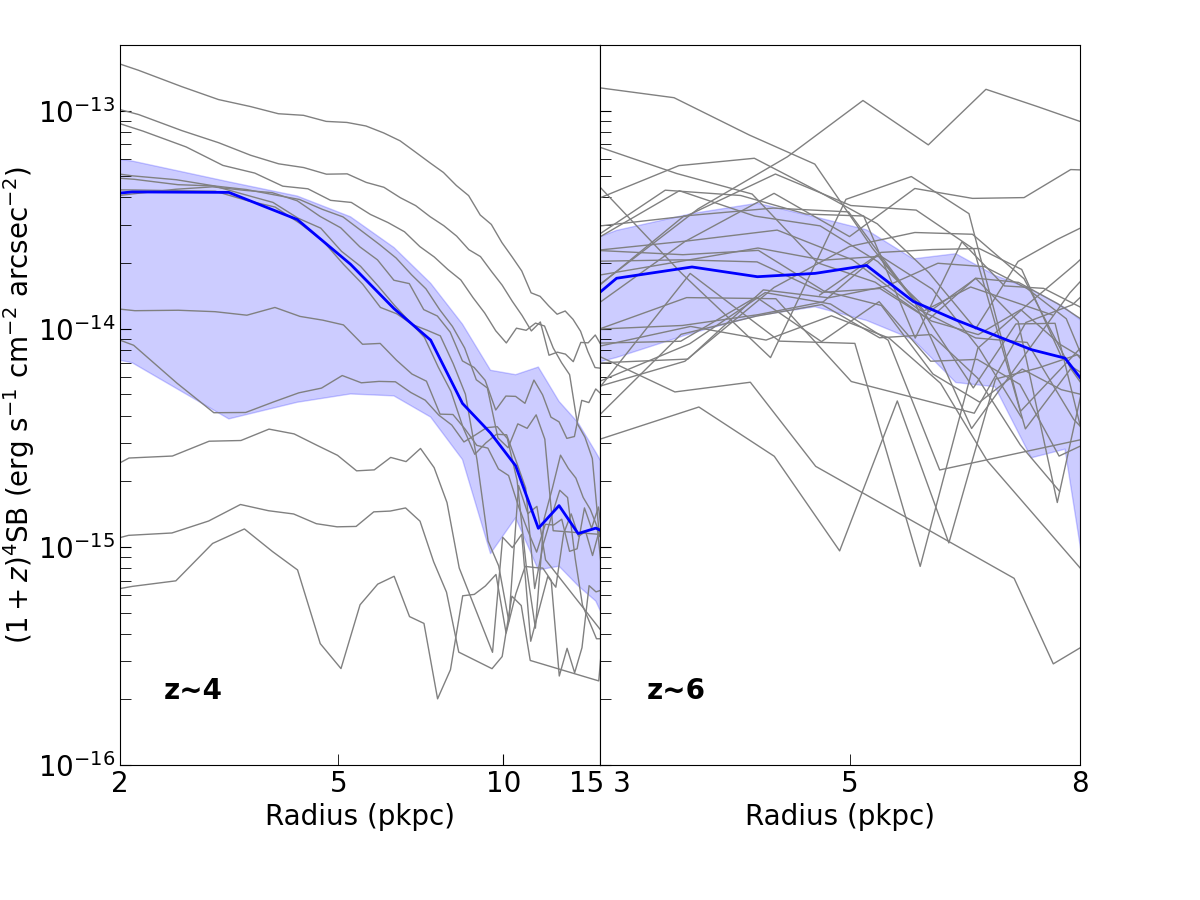}
    \caption{\Lya surface brightness profiles at $z\sim4$ (left) and $z\sim6$ (right) quasar samples. Median profiles are also shown as the blue lines. The shaded regions represent the 25th--75th percentiles of the variation in each quasar sample.}
    \label{fig:each_profile}
\end{figure}

In this section, we discuss the SB profiles of the \Lya halo. 
The \Lya flux at each radius is derived by integrating the residual fluxes in the wavelength range, where the SNR is greater than two for $z\sim4$, and between $-500$ and $+500\ \mathrm{km\ s^{-1}}$ for $z\sim 6$, and adding together the flux at the same radius from the centre in the slit direction.
The corresponding area is calculated by the slit width ($=1\arcsec$) $\times$ pixel scale ($=0.1185\arcsec,0.2076\arcsec, \rm{and}\ 0.254\arcsec$ for DEIMOS, FOCAS, and OSIRIS). The profiles are smoothed by taking a moving average over four ($\sim 3.4$ pkpc, 4.7 pkpc, and 5.7 pkpc for DEIMOS, FOCAS, and OSIRIS) bins, and the cosmological dimming is corrected by multiplying $(1+z)^4$. 
The radial SB profile of each object is adjusted to the physical scale corresponding to its redshift, and the median of the radial profile is taken for each sample at $z\sim4$ and $z\sim6$. 
The individual SB profiles, which show a large variation, are shown in Figure~\ref{fig:each_profile} together with the median SB profile. 
Although we can measure the integral \Lya halo flux in our sample (Section~\ref{sec:ldep}), their individual radial SB profiles are difficult to discuss due to the lack of sufficient SNR. In the following, we discuss the median profile, which is made by median stacking of all radial profiles of \Lya halo detected objects.
Since the \Lya halo is spatially asymmetric, it is inevitable that the slit direction can easily affect the measurements of the radial SB profile and \Lya halo luminosity. 
However, assuming that the slit direction is random with respect to the spreading direction of quasar \Lya halo, the median measurements based on a certain number of samples is likely to be less affected.

\subsubsection{SB profiles at $z\sim4$}
\begin{figure}
\includegraphics[width=\columnwidth]{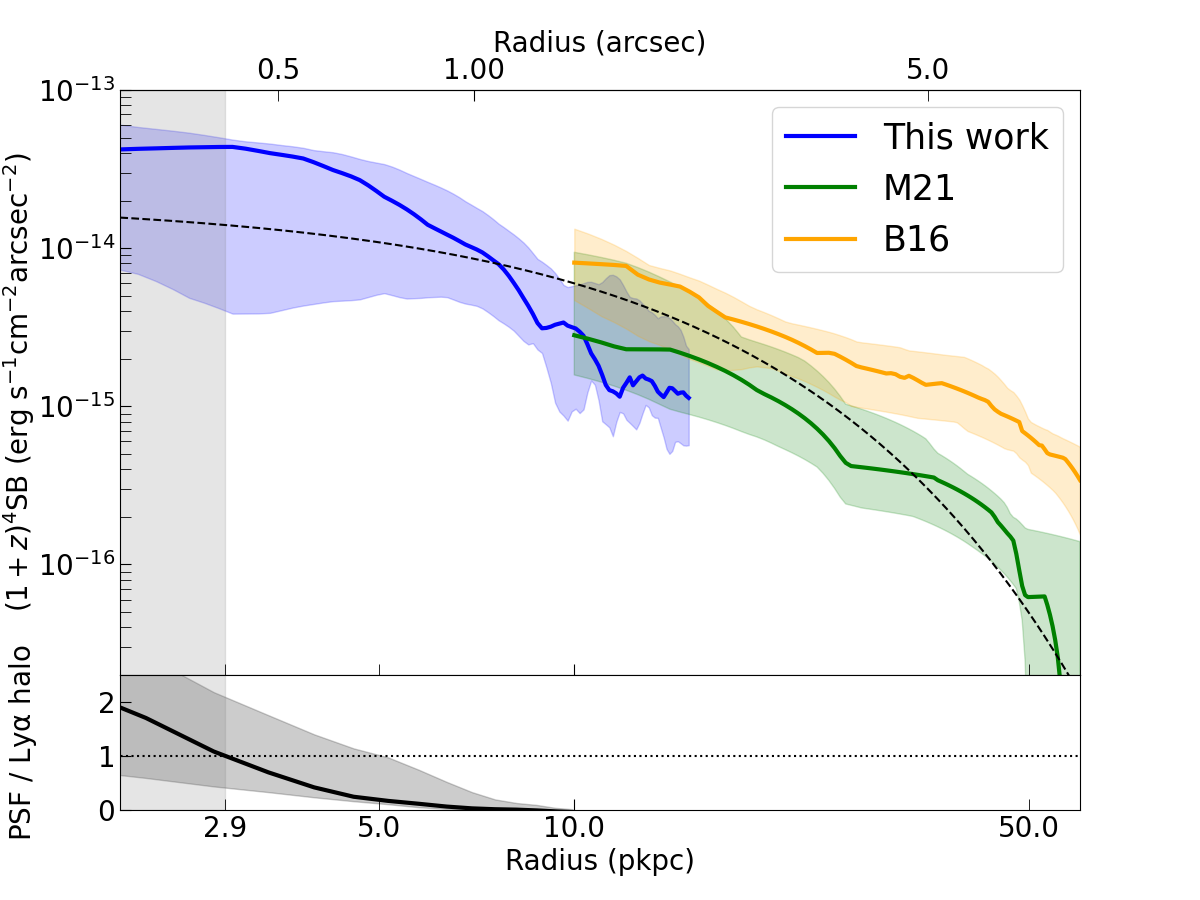}
    \caption{Median \Lya surface brightness profiles at $z\sim4$ quasar sample (blue). 
    The median profile of \citetalias{mackenzie2021revealing} is shown as a green line, and \citetalias{borisova2016ubiquitous} is shown as an orange line.
    Shaded regions represent the 25th--75th percentiles of the variation in each quasar sample.
    The best-fit exponential fit is also shown as a dashed line. The upper horizontal axis shows the spatial radius in the unit of arcsec, assuming at $z=3.71$. The lower panel shows the ratio of the median PSF SB profile to the median \Lya halo SB profile. Uncertainty is evaluated using the 25th--75th percentiles of the variation of the PSF profile. The region where PSF/\Lya halo $> 1$ is shaded in grey.}
    \label{fig:radial4}
\end{figure}

Figure~\ref{fig:radial4} shows the median SB profile of the \Lya halo at $z\sim4$. 
Near the centre, the PSF dominates, leading to greater uncertainties in the PSF subtraction. 
Therefore, we exclude such central regions from the analysis by comparing the median profiles of the PSF and the \Lya halo. 
The ratio of the median PSF profile to the median \Lya SB profile is shown in the lower panel of Figure~\ref{fig:radial4}. 
The uncertainty in the median PSF profile is estimated in the same way as the median \Lya profile, using the 25th--75th percentiles of the variation in the PSF profile. This ratio decreases towards the outer regions and the PSF variation also decreases with radius.
We do not use the inner 2.9 pkpc ($\sim 0.41\arcsec$) of the profile, where the median \Lya SB is smaller than the median PSF SB (grey shaded in Figure~\ref{fig:radial4}), in our analysis. 
Our slit spectroscopy can determine the radial profile more inward than that obtained with MUSE, which is difficult due to its complicated three-dimensional PSF profile.
Conversely, the SNR decreases rapidly on the outside ($> 15$ pkpc), where it is hard to measure the profile.
This is a weakness of the slit spectroscopy, which, unlike IFU, cannot take a sufficiently large area on the outside.

Our profile matches well that of \citetalias{mackenzie2021revealing}, a sample with quasars of similar brightness (median $M_{1450}=-24.40$) to ours (median $M_{1450}=-23.59$), at $r\sim 10$ pkpc, while the profile of \citetalias{borisova2016ubiquitous}, a brighter sample, is $\sim 0.5$ dex brighter than ours. 
This is qualitatively consistent with \citetalias{mackenzie2021revealing}, showing that the fainter quasars host fainter haloes. We will discuss the luminosity dependence in Section~\ref{sec:ldep}.
Radial SB profile is known to be well fitted by an exponential function~\citep[e.g.][]{battaia2018,farina2019requiem}: $(1+z)^4 \mathrm{SB} = C\exp{(-r/r_h)}$, where $C$ is the normalization factor and $r_h$ is the scale length of the profile. At $z\sim4$, an exponential fit is performed to the combined radial profile of ours and \citetalias{mackenzie2021revealing}, joined at $r=10\ \mathrm{pkpc}$. 
The best-fit curve with parameters, $C=(2.61\pm 0.13)\times 10^{-14}\ \mathrm{erg\ s^{-1} cm^{-2} arcsec^{-2}}$ and $r_h = 7.73\pm 0.14$ pkpc, is shown in Figure~\ref{fig:radial4}. 
The best-fit exponential curve is within the 25th--75th percentile range, and it well represents the profile. 

\subsubsection{SB profiles at $z\sim6$}
\begin{figure}	\includegraphics[width=\columnwidth]{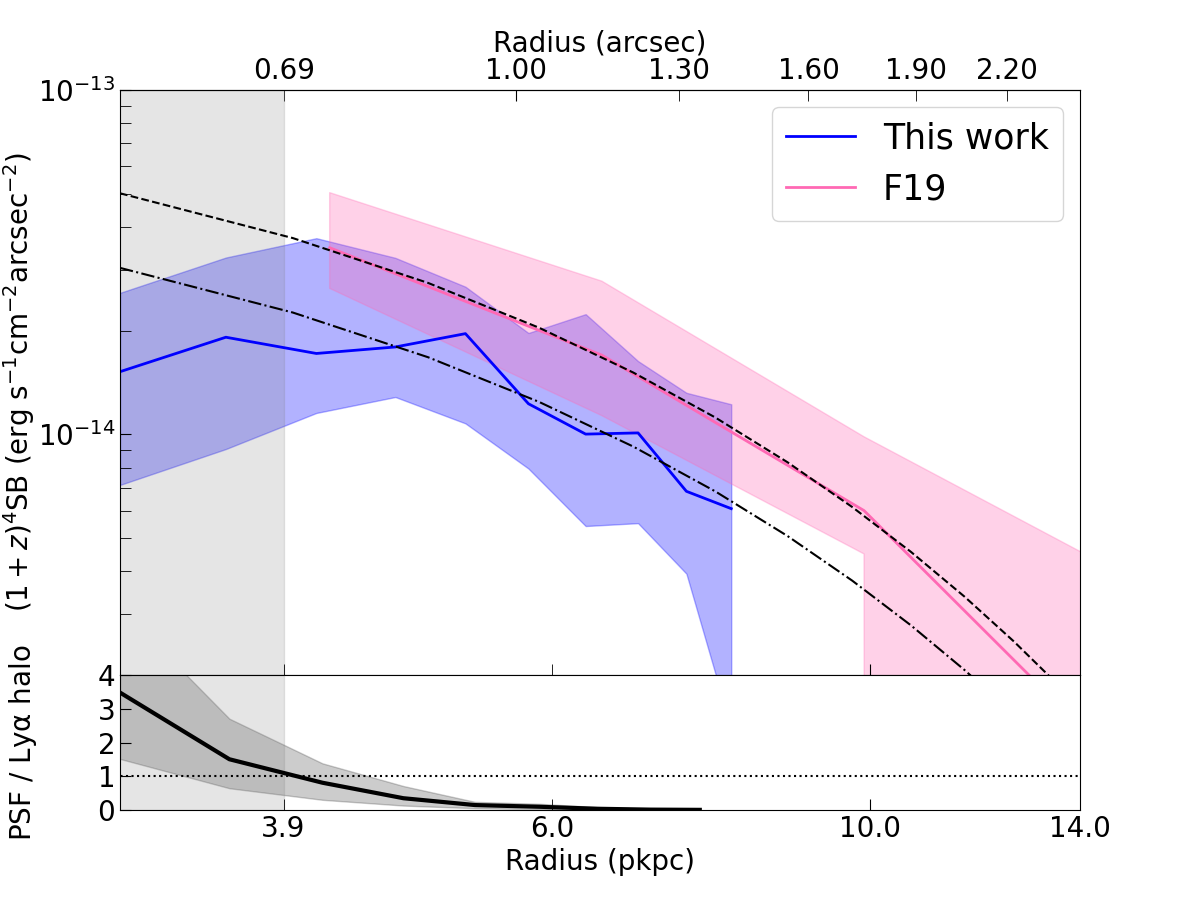}
    \caption{Median \Lya surface brightness profiles around $z\sim6$ quasar sample (blue), compared with that of \citetalias{farina2019requiem} (pink). Shaded regions represent the 25th--75th percentiles of the variation in each quasar sample. 
    The best-fit exponential fit for \citetalias{farina2019requiem} is shown as a dashed line, and the best fit for our sample, fixed to the same $r_h$ as \citetalias{farina2019requiem}, is also shown in the dashed-dotted line. The upper horizontal axis corresponds to the spatial radius in the unit of arcsec, assuming at $z=6.09$. 
    In the lower panel, the ratio of the median PSF SB profile to the median \Lya SB profile and the region where PSF/\Lya halo $> 1$ are shown as in Figure~\ref{fig:radial4}.
    }
    \label{fig:radial6}
\end{figure}

Figure~\ref{fig:radial6} shows the median SB profile of the \Lya halo at $z\sim6$.
The inner 3.9 pkpc ($\sim 0.69\arcsec$) of the profile, the grey-shaded region in Figure~\ref{fig:radial6}, is not used in our analysis due to the large PSF uncertainty. 
The inner radius is larger than that at $z\sim4$, which is mainly due to the significant SB reduction of \Lya halo by cosmic dimming at $z\sim6$, and partly due to the use of a single Gaussian to estimate the PSF at $z\sim6$, resulting in a tendency to overestimate the PSF amplitude. 
The \citetalias{farina2019requiem} profiles are provided from $r=4.2$ pkpc to the outside, well beyond our measurement.
Note that the median profile of \citetalias{farina2019requiem} presented here is constructed by stacking only their detected samples to compare with our analysis consistently.

The amplitude of our median profile is systematically lower than that of \citetalias{farina2019requiem}, showing that the fainter quasars host fainter haloes. This result is consistent with our result at $z\sim4$ but inconsistent with \citetalias{farina2019requiem}.
Median SB profile of \citetalias{farina2019requiem} is well fitted by the exponential curve with the parameters, $C=(1.28\pm 0.01)\times 10^{-13}\ \mathrm{erg\ s^{-1} cm^{-2} arcsec^{-2}}$ and $r_h = 3.21\pm 0.02$ pkpc. If we fix $r_h = 3.21$ pkpc and fit our profile with only $C$ as a free parameter, $C$ is determined to be $C=(7.75\pm 0.60)\times 10^{-14}\ \mathrm{erg\ s^{-1} cm^{-2} arcsec^{-2}}$, which is smaller than that of \citetalias{farina2019requiem}.


\subsection{Luminosity dependence} \label{sec:ldep}
We investigate the correlation between the luminosity of \Lya haloes and the quasar luminosity in more detail.
Assuming that the profile of all halos can be approximated by an exponential function, we calculate $C$ and $r_h$ for each halo within the observed radius range.
Then, the total luminosity for each halo, $L_\mathrm{halo}$, is estimated by integrating the resulting best-fit exponential function up to $r=100$ pkpc, sufficiently outside each halo.
Even if we change this 100 pkpc slightly, the estimate of $L_\mathrm{halo}$ is hardly changed.
The typical $r_h$ of our sample are 7.73 pkpc and 3.21 pkpc at $z\sim4$ and $z\sim6$, and when $r/r_h$ exceeds five, the increase in $L_\mathrm{halo}$ is only within 1\%. 
A few samples that significantly deviate from the exponential function are removed using $3\sigma$-clipping for $C$ and $r_h$. 
The error of the $L_\mathrm{halo}$ is evaluated from the background variance and the fitting error.
As seen in the previous section, the radii traced in the study differ from those in previous studies.
For $z\sim4$, our sample is fitted in the range of $2.9 < r < 15\ \mathrm{pkpc}$, while \citetalias{mackenzie2021revealing} and \citetalias{borisova2016ubiquitous} are fitted in the range of $10 < r < 60\ \mathrm{pkpc}$. 
For $z\sim6$, our sample is fitted in the range of $3.9 < r < 8\ \mathrm{pkpc}$, while \citetalias{farina2019requiem} is fitted using all available data points. 

We also evaluate the upper limits of $L_\mathrm{halo}$ for non-detected objects.
It is challenging because our $L_\mathrm{halo}$ estimates are based on the flux measurements within 2$\arcsec$ and extend the profile outward, whereas the actual data is limited not only one-dimensional on a slit, but also does not extend far enough outward, making it difficult to accurately estimate the residual flux over the halo area. 
Therefore, at first, the flux residuals within 2$\arcsec$ from the centre on the two-dimensional spectrum is measured for non-detected objects, and then extrapolating it to the area of the halo to estimate the upper limit of $L_\mathrm{halo}$, assuming the flux residuals are almost the same outside the real data. 
This extrapolation is a large assumption, and may not accurately evaluate the upper limit, especially when the halo is asymmetric, but we believe this is the best effort. 

We use a couple of indicators for quasar luminosity: the luminosity at the Lyman limit, $L_\mathrm{{\nu_{LL}}}$, and the luminosity at the spectral peak of the \Lya line, $L\mathrm{^{peak}_{Ly\alpha;QSO}}$~\citep{2019MNRAS.482.3162A}. 
The magnitude, $M_{912}$ corresponding to 
$L_\mathrm{{\nu_{LL}}}$ can be derived from the $M_{1450}$ using
\begin{equation}
    M_{912} = M_{1450}+0.33 \label{eq:M14502M912}.
\end{equation}

For $z\sim4$ sample, $L\mathrm{^{peak}_{Ly\alpha;QSO}}$ is given in \citetalias{mackenzie2021revealing} and \citetalias{borisova2016ubiquitous}, and the $L\mathrm{^{peak}_{Ly\alpha;QSO}}$ of our sample is measured from the quasar one-dimensional spectrum, which is made by integrating the flux within 2$\arcsec$ radius aperture and smoothed to match the wavelength resolution of MUSE. 
The $L^{\mathrm{peak}}_{\mathrm{Ly\alpha;QSO}}$ for our $z\sim6$ sample is measured in the same way as at $z\sim4$, following \citetalias{mackenzie2021revealing} and \citetalias{borisova2016ubiquitous}.
The relation is also empirically determined by \citet{2015MNRAS.449.4204L} as in Equation (\ref{eq:Mi2M1450}).


We should note that $L\mathrm{_{Ly\alpha;QSO}^{peak}}$ might not be an accurate indicator of the quasar \Lya luminosity at $z\sim6$, because it is more or less affected by absorption by neutral hydrogen in IGM. 
For this reason, $L\mathrm{_{Ly\alpha;QSO}^{peak}}$ is not measured in \citetalias{farina2019requiem}, so we discuss the dependence of the $L_\mathrm{{halo}}$ on $L^{\mathrm{peak}}_{\mathrm{Ly\alpha;QSO}}$ at $z\sim6$ using only our sample.
We note that our $L\mathrm{_{Ly\alpha;QSO}^{peak}}$ - $\ M_{1450}$ relation at $z\sim6$ follows the distribution at $z<4$ (e.g. Figure 1 in \citetalias{mackenzie2021revealing}), inferring that IGM attenuation is not significant. 

\begin{figure}	\includegraphics[width=\columnwidth]{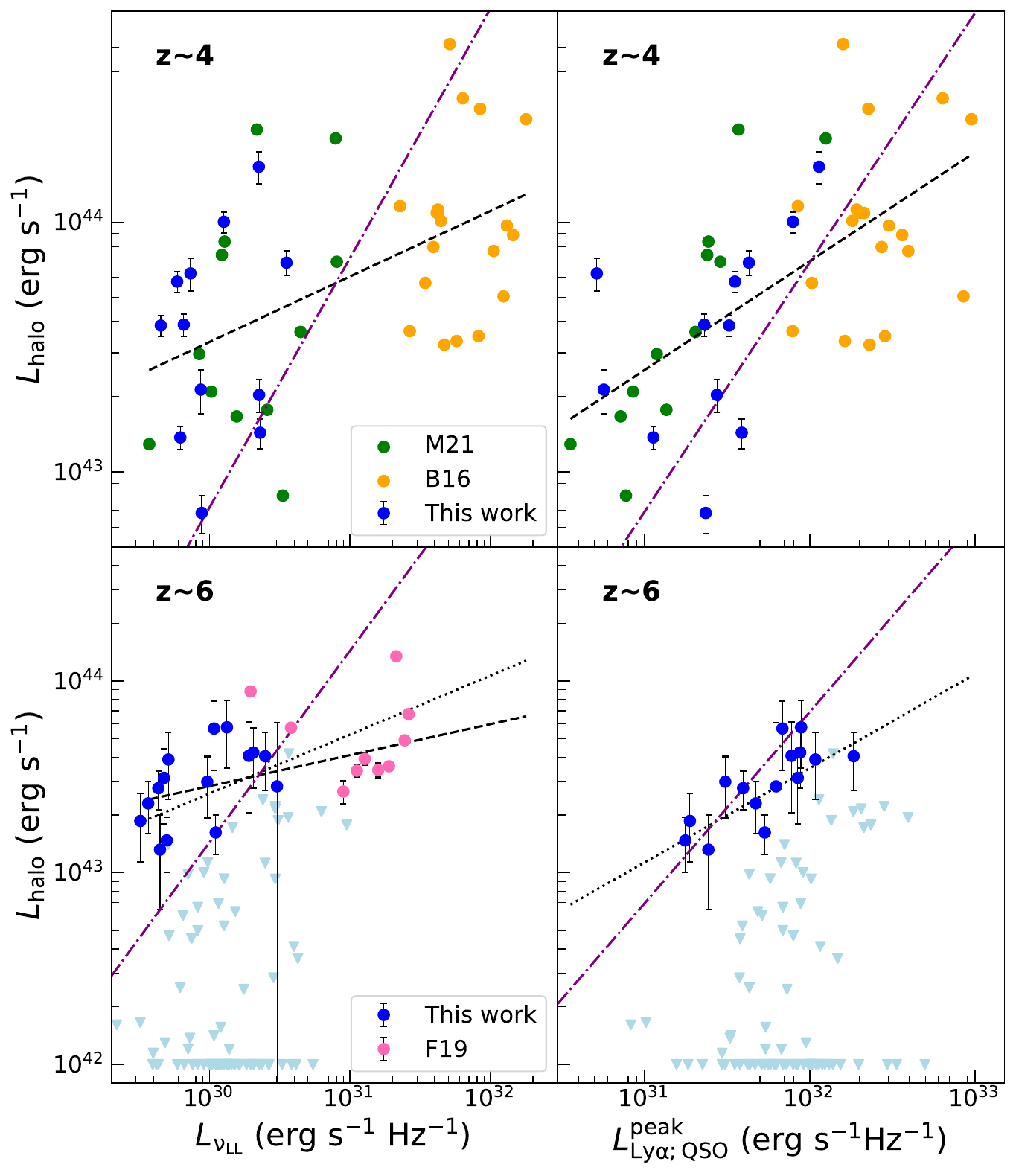}
    \caption{The luminosity dependence of $z\sim4$ (top) and $z\sim6$ (bottom) samples. \Lya halo luminosities are plotted against both the quasar luminosity at the Lyman limit, $L_\mathrm{{\nu_{LL}}}$ (left) and 
    at the peak of the \Lya line (right), $L^{\mathrm{peak}}_{\mathrm{Ly\alpha;QSO}}$. Our quasars are shown as blue circles with \citetalias{borisova2016ubiquitous} (orange) and \citetalias{mackenzie2021revealing} (green) for $z\sim4$ and \citetalias{farina2019requiem} (pink) for $z\sim6$. The linear best fits including the brighter sample are also shown as the dashed line. The best fit for only our $z\sim6$ sample is shown as the dotted line. The purple dashed-dotted line represents the expected relation derived from Equations~\ref{thickL}(left) and~\ref{eq:scatL}(right). We set $f_c = 0.01$ and $0.02$ (see Section~\ref{sec:thick_rec}) at $z\sim4$ and $z\sim6$ respectively in Equation~\ref{thickL}, and $f_c = 0.5$ in Equation~\ref{eq:scatL}. The upper limits of $L_\mathrm{{halo}}$ for non-detected objects are shown as light blue triangles at $z\sim6$. Objects with upper limit $L_\mathrm{halo}<1.0\times10^{42}\ \mathrm{erg\ s^{-1}}$ are shown at $L_\mathrm{halo}=1.0\times10^{42}\ \mathrm{erg\ s^{-1}}$
    }
    \label{fig:Ldep}
\end{figure}

\begin{table*}
	\centering
	\caption{The results of the Spearman's rank correlation test and the linear fit to Figure~\ref{fig:Ldep}}
	\label{tab:parameters}
\begin{tabular}{lcccccc}
        \hline
         & & $L_\mathrm{{\nu_{LL}}}$ &  &  & $L^{\mathrm{peak}}_{\mathrm{Ly\alpha;QSO}}$ & \\
		\hline
        Sample & r & p & Slope & r & p & Slope\\
		\hline
		$z\sim4$ This work+\citetalias{mackenzie2021revealing}+\citetalias{borisova2016ubiquitous} & 0.49 & $8.2\times 10^{-4}$ & 0.26$\pm$0.06 & 0.63 & $5.7\times 10^{-6}$ & 0.44$\pm$0.08\\
		\hline
        $z\sim6$ This work+\citetalias{farina2019requiem} & 0.55 & $5.3\times 10^{-3}$ & 0.15$\pm$0.06 & - & - & - \\
		\hline
        $z\sim6$ This work & 0.62 & $1.9\times 10^{-2}$ & 0.34$\pm$0.19 & 0.69 & $6.5\times 10^{-3}$ & 0.48$\pm$0.13\\
		\hline
	\end{tabular}\\
\end{table*}

Figure~\ref{fig:Ldep} shows the $L_\mathrm{halo}$ dependencies on quasar luminosities. We discuss the significance of dependencies using Spearman's rank correlation test. The Spearman's rank correlation coefficients, r, and p-values are listed in Table~\ref{tab:parameters}. At $z\sim4$, the p-values are $8.2\times10^{-4}$ and $5.7\times10^{-6}$ for $L_\mathrm{{\nu_{LL}}}$ and $L^{\mathrm{peak}}_{\mathrm{Ly\alpha;QSO}}$, respectively, both suggesting a statistically robust correlation.
The Spearman's rank correlation coefficients are 0.49 and 0.63, suggesting that the correlation with $L^{\mathrm{peak}}_{\mathrm{Ly\alpha;QSO}}$ is slightly stronger, though, since $L_\mathrm{{\nu_{LL}}}$ and $L^{\mathrm{peak}}_{\mathrm{Ly\alpha;QSO}}$ are correlated with each other, it is difficult to conclude which is the essential correlation. 
The result that $L_\mathrm{halo}$ is correlated with both $L_\mathrm{{\nu_{LL}}}$ and $L^{\mathrm{peak}}_{\mathrm{Ly\alpha;QSO}}$ is consistent with \citetalias{mackenzie2021revealing}.
The luminosity dependence is also found at $z\sim 6$ as shown in the bottom panels of Figure~\ref{fig:Ldep}. 
The p-values based on our sample only are $1.9\times 10^{-2}$ and $6.5\times 10^{-3}$, which are sufficiently small, and $r$ are 0.62 and 0.69, for $L_\mathrm{{\nu_{LL}}}$ and $L^{\mathrm{peak}}_{\mathrm{Ly\alpha;QSO}}$, respectively.
The $L_\mathrm{halo}$ is correlated with both $L_\mathrm{{\nu_{LL}}}$ and $L^{\mathrm{peak}}_{\mathrm{Ly\alpha;QSO}}$, and the correlation is slightly stronger with $L^{\mathrm{peak}}_{\mathrm{Ly\alpha;QSO}}$. 
When including the \citetalias{farina2019requiem} sample in the $L_\mathrm{{\nu_{LL}}}$ relation, the p-value is $5.3\times 10^{-3}$, indicating a significant dependence on $L_\mathrm{{\nu_{LL}}}$. Comparing the strength of the correlations with $L_\mathrm{{\nu_{LL}}}$, for which bright samples are available both at $z\sim4$ and $z\sim6$, we find that the $r$ is slightly higher at $z\sim6$, indicating tighter correlations at $z\sim6$. 
The slopes of the linear best fit for Figure~\ref{fig:Ldep} are listed in Table~\ref{tab:parameters}. 
Interestingly, the slopes at $z\sim4$ and $z\sim6$ are consistent within the error, inferring a similar physical mechanism producing \Lya halo.
However, as discussed in Section~\ref{sec:detection}, the detection of haloes in our sample is incomplete, and it is possible that we have not detected haloes with smaller $L_\mathrm{halo}$ even if the quasar luminosities are similar. 
Even in this case, the correlation remains, but it should be noted that we have not obtained strong constraints on the strength or slope of the correlation.
We conclude that there is a quasar luminosity (both $L_\mathrm{{\nu_{LL}}}$ and $L^{\mathrm{peak}}_{\mathrm{Ly\alpha;QSO}}$) dependence in $L_\mathrm{halo}$ at $z\sim 6$ as well as $z\sim 4$.
\citetalias{farina2019requiem} find no dependence on quasar luminosity probably due to their narrow luminosity range, while the result in this study is achieved by expanding the luminosity dynamic range.

\subsection{Correlation with central BH} \label{sec:result_bh}

We also assess a possible dependence of $L_\mathrm{halo}$ on the BH mass ($M\mathrm{_{BH}}$) and the Eddington ratio ($\lambda\mathrm{_{edd}}$). 
We retrieve the $M\mathrm{_{BH}}$ and $\lambda\mathrm{_{edd}}$ from \citet{he2024black} for all of our faint $z\sim4$ sample.
Also, the $M\mathrm{_{BH}}$ and $\lambda\mathrm{_{edd}}$ of ten quasars in  \citetalias{mackenzie2021revealing} sample are found in \citet{wu2022}.
Although we also look for the $M\mathrm{_{BH}}$ and $\lambda\mathrm{_{edd}}$ measurements for \citetalias{borisova2016ubiquitous} sample, we find only two measurements in \citet{wu2022}.
For $z\sim6$ samples, we use available measurements for eight quasars by \citet{farina2022} for \citetalias{farina2019requiem}.
All of these studies estimate $M\mathrm{_{BH}}$ using the single-epoch method with the C{\sc iv} emission line, given by \citet{vestergaard2006}. 
We use \citet{takahashi2024} for 11 of our $z\sim6$ faint sample; however, they estimate the $M\mathrm{_{BH}}$ for each $z\sim6$ quasar by combining the measured continuum magnitude and the C{\sc iv} line width of a low-$z$ SDSS quasar whose spectrum at the rest-frame $\sim1200-1400$ \AA$\ $match with that of the quasar.
Because their method is not usual, the reliability of the $M\mathrm{_{BH}}$ obtained from it is different from that of other samples based on conservative measurements.
The bolometric luminosity to estimate the $\lambda\mathrm{_{edd}}$ 
is calculated from the luminosity at 3000 \AA $\ $ using the relation derived by \citet{richards2006sed} and updated by \citet{shen2011} in \citet{farina2022}, while for the others, it is calculated from the luminosity at 1350 \AA $\ $ using the relation derived by \citet{richards2006sed}.

\begin{figure}	\includegraphics[width=\columnwidth]{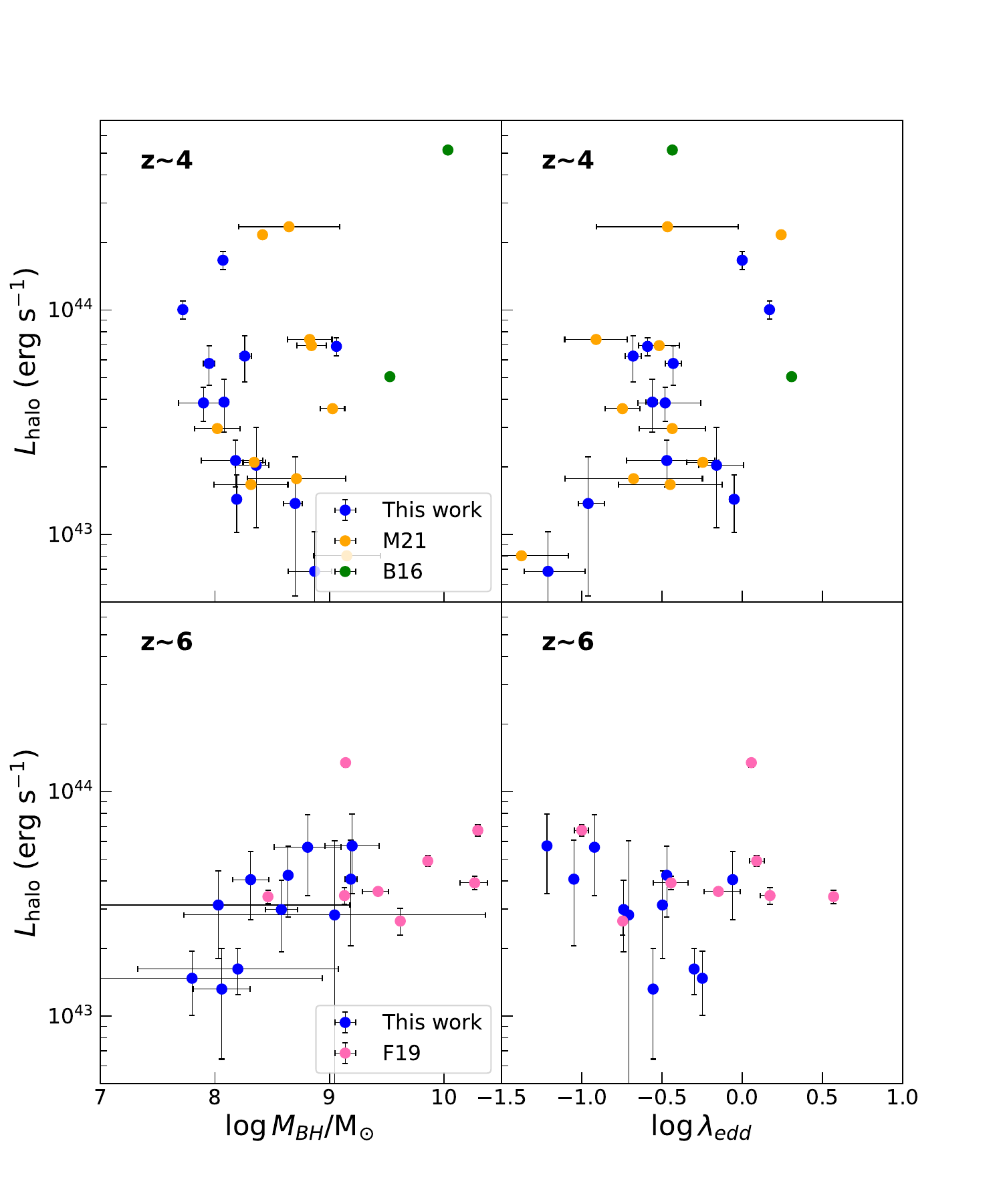}
    \caption{The $L\mathrm{_{halo}}$ dependencies on $M\mathrm{_{BH}}$ (left) and $\lambda\mathrm{_{edd}}$ (right) at $z\sim4$ (top) and $z\sim6$ (bottom). Our quasars are shown as blue circles with \citetalias{borisova2016ubiquitous} (orange) and \citetalias{mackenzie2021revealing} (green) for $z\sim4$ and \citetalias{farina2019requiem} (pink) for $z\sim6$.
    }
    \label{fig:bhmass}
\end{figure}

\begin{table*}
	\centering
	\caption{The results of the Spearman's rank correlation test on the relation shown in Figure~\ref{fig:bhmass}.}
	\label{tab:bhmass}
\begin{tabular}{rcc}
		\hline
         & r & p \\
		\hline
		$z\sim4$ $M_{\mathrm{BH}}$ & $-0.04$ & $8.6\times 10^{-1}$ \\
        \hline
		$\ \lambda\mathrm{_{edd}}$ & 0.39 & $6.3\times 10^{-2}$ \\
        \hline
		$z\sim6$ $M_{\mathrm{BH}}$ & 0.58 & $9.4\times 10^{-3}$ \\
        \hline
		$\ \lambda\mathrm{_{edd}}$ & $-0.11$ & $6.6\times 10^{-1}$ \\
		\hline
	\end{tabular}\\
\end{table*}

Figure~\ref{fig:bhmass} shows the $L\mathrm{_{halo}}$ dependencies on $M\mathrm{_{BH}}$ and $\lambda\mathrm{_{edd}}$, and the Spearman's rank correlation coefficients, r, and p-values are listed in Table~\ref{tab:bhmass}. A statistically robust correlation (i.e. $p<0.05$) is found only between $M\mathrm{_{BH}}$ and $L\mathrm{_{halo}}$ at $z\sim6$, while not for the others. 
At $z\sim6$, the dependence of $L\mathrm{_{halo}}$ on $M\mathrm{_{BH}}$, which had not been found by \citet{farina2022}, is confirmed by widening the range of $M\mathrm{_{BH}}$.
At $z\sim4$, on the other hand, there is no correlation with $M\mathrm{_{BH}}$, and there is a weak correlation with $\lambda\mathrm{_{edd}}$, although this is not statistically significant.
We should note that the available data in \citetalias{borisova2016ubiquitous}, the bright sample at $z\sim4$, are limited to only two, which might be too small to see the correlation with the dynamic range of $M\mathrm{_{BH}}$ fully expanded. 
The lack of robust correlation at $z\sim4$ might be due to an insufficient number of bright samples. 
Similarly, the measurement of $M\mathrm{_{BH}}$ for faint quasars at $z\sim6$ is not based on a conservative method, thus it is not possible to completely rule out the possibility that it is causing some bias.
To accurately investigate the correlations, it is necessary to increase the sample size across a wide range of parameter space, with both $M\mathrm{_{BH}}$ and $L\mathrm{_{halo}}$ available.

\section{DISCUSSION}\label{sec:discussion}
In this section, we discuss several powering mechanisms of the \Lya haloes. In general, three mechanisms \citep[e.g.][]{henwai2013,hennawi2015,2019MNRAS.482.3162A,cantalupo2017} are thought to contribute to the emission from \Lya halo: recombination (also called \Lya fluorescence), scattering of \Lya photons produced by the quasar broad line region, and collisional excitation. 
The mechanisms that work most effectively depend on the quasar luminosity as well as the physical state of the CGM, such as density, temperature and ionization state. 
From an observational perspective, quasar luminosity dependence on the \Lya halo luminosity is one of the effective observational quantities that can constrain the emission mechanism.

\subsection{Recombination radiation}\label{sec:recombination}
First, we discuss the contribution of recombination radiation following a similar approach to \citetalias{farina2019requiem}. \Lya photons are produced in a photoionized medium as a result of a recombination cascade. Assuming that the quasar is surrounded by cold gas clouds, the surface brightness of the emission can be estimated under two scenarios: optically thick ($N_{\rm{H_I}} \gg 10^{17.2}\ \rm{cm}^{-2}$) and thin ($N_{\rm{H_I}} \ll 10^{17.2}\ \rm{cm}^{-2}$).

\subsubsection{Optically thick scenario}\label{sec:thick_rec}
In this case, the quasar radiation is self-shielded, and \Lya emission originates from a thin, highly ionized envelope around an individual cloud. Given that there is sufficient gas to be ionized, it is expected that as the ionizing radiation from the quasar increases, the total emission from the gas clouds also increases. 

\citet{henwai2013} and \citet{hennawi2015} provide a prescription of observed Ly$\alpha$ SB when the central quasar is surrounded by cold optically thick clouds uniformly
spatially distributed in a spherical halo of radius $R$:
\begin{multline}
    \mathrm{SB} = {\frac{\eta_\mathrm{thick}h\nu_\mathrm{Ly\alpha}}{4\pi (1+z)^4}} f_c \Phi (R/\sqrt{3})\\
    \hspace{0.4cm}= 2.3\times10^{-15} (1+z)^{-4} \left(\frac{f_c}{0.5}\right)\left(\frac{R}{100\ \mathrm{kpc}}\right)^{-2}\\
    \times \left(\frac{L_\mathrm{\nu_{LL}}}{10^{30}\ \mathrm{erg\ s^{-1}Hz^{-1}}}\right)\ \mathrm{erg \ s^{-1}cm^{-2}arcsec^{-2}}\label{eq:thickSB}
\end{multline}
where $\eta_{\rm thick}$ is a
fraction of incident photons converted into \Lya by the cloud's envelope, $f_c$ is a covering fraction. $\Phi$ is the ionizing photon number flux given by
\begin{equation}
    \Phi(r) = \int^{\infty}_{\nu_\mathrm{LL}} \frac{F_\nu}{h\nu}d\nu 
    =  \frac{L\mathrm{_{\nu_{LL}}}}{4\pi r^2}\int^{\infty}_{\nu_\mathrm{LL}} \frac{1}{h\nu}\left(\frac{\nu}{\nu_\mathrm{LL}}\right)^{\alpha_{\mathrm{UV}}} d\nu
\end{equation}
where $\nu_\mathrm{LL}$ is the frequency at the Lyman limit, and we assume that quasar spectral energy distribution follows the power law with the slope of $\alpha_\mathrm{UV},\ L_\nu = L_{\nu_\mathrm{LL}}(\nu/\nu_\mathrm{LL})^{\alpha_\mathrm{UV}}$ for blueward of $\nu_{\mathrm{LL}}$. We also assume $\eta_\mathrm{thick}=0.66$ based on the theoretical calculations of \citet{gould1996} and $\alpha_\mathrm{UV}=-1.7$ based on the measurement of \citet{2015MNRAS.449.4204L} as with \citet{hennawi2015}.

Following the \citet{henwai2013}, the total \Lya luminosity of \Lya halo is given by
\begin{multline}
   L_\mathrm{halo} = 4\pi^2(1+z)^4 R^2 \mathrm{SB}\\
   \hspace{0.68cm}= 1.6\times10^{45}\left(\frac{R}{100\ \mathrm{kpc}}\right)^2 \\ \times (1+z)^4 \left(\frac{\mathrm{SB}}{10^{-14}\ \mathrm{erg \ s^{-1}cm^{-2}arcsec^{-2}}}\right)\ \mathrm{erg\ s^{-1}}
   \label{eq:sb2l}.
\end{multline}
From Equations (\ref{eq:thickSB}) and (\ref{eq:sb2l}), we can write
\begin{equation}
    L_\mathrm{halo} = 3.6\times 10^{44} \left(\frac{f_c}{0.5}\right)\left(\frac{L_\mathrm{\nu_{LL}}}{10^{30}\ \mathrm{erg\ s^{-1}Hz^{-1}}}\right)\ \mathrm{erg\ s^{-1}}.
    \label{thickL}
\end{equation}

Thus in the optically thick scenario, the \Lya halo luminosity should be proportional to $L\mathrm{_{\nu_{LL}}}$.
As discussed in Section~\ref{sec:ldep}, they show linearly correlated at both $z\sim 4$ and $z\sim6$, but the relations are shallower than proportional,  which is shown as the purple dashed-dotted line in Figure~\ref{fig:Ldep}.
This is likely to be due to the fact that the quasar CGMs are not perfectly optically thick, as discussed below. 
Therefore, recombination radiation from optically thick regions may partly contribute to the emission.
It also should be noted that the dependence on $L_\mathrm{\nu_{LL}}$ does not necessarily rely on recombination radiation from optically thick gas. 
There is also a correlation between $L_\mathrm{\nu_{LL}}$ and $L\mathrm{^{peak}_{Ly\alpha;QSO}}$ (see e.g. Figure 1 in \citetalias{mackenzie2021revealing}), and if there is a correlation between the \Lya halo luminosity and $L^{\mathrm{peak}}_{\mathrm{Ly\alpha;QSO}}$, which is actually stronger than with $L_\mathrm{\nu_{LL}}$ (see Figure~\ref{fig:Ldep}), it may appear that there is also an indirect correlation with $L_\mathrm{\nu_{LL}}$. 

To investigate this in more detail, we estimate the $L\mathrm{_{halo}}$ numerically using Equation (\ref{thickL}). 
According to \citet{battaia2015a,battaia2015b} and \citet{hennawi2015}, $f_c \ga 0.5$ is implied by the smooth morphology of the emission in the \Lya halo. 
Given that the median luminosities of our sample at
the Lyman edge are $L\mathrm{_{\nu_{LL}}} = 8.7\times 10^{29}, 1.0\times 10^{30}\  \mathrm{erg\ s^{-1} Hz^{-1}}$ for $z\sim4,6$, median \Lya halo luminosities are estimated to be $L_\mathrm{halo}\ga 3.1\times 10^{44}, 3.6\times 10^{44}\ \mathrm{erg\ s^{-1}}$, respectively.
These are about one dex larger than those observed, leading to a discrepancy with the observations if we assume that the gas is completely optically thick. From the above, we conclude that the recombination radiation from optically thick gas, presumably in the H{\sc i} rich region near the centre, is likely to contribute, but that the \Lya halo does not light up solely by this radiation.

If only a limited region around the centre is optically thick in the halo,  the covering fraction might be much lower. 
The observed median \Lya halo luminosities, $L_\mathrm{halo} = 3.9\times10^{43},\ 3.5\times10^{43}\ \mathrm{erg\ s^{-1}}$, are reproduced by Equation~(\ref{thickL}) if we assume $f_c=0.06,\ 0.05$ at $z\sim4$ and $z\sim6$. 
If we include the bright sample, the observed \Lya halo luminosities are roughly consistent with the assumption of $f_c\sim0.01,\ 0.02$, respectively, as shown by the purple dashed-dotted lines in Figure~\ref{fig:Ldep}. 
Note that this is a very roughly determined $f_c$ with an analytical formula to reproduce the observed median \Lya halo luminosity. 
Such a significant low covering fraction of $<0.1$ is also found at $z\sim2$ by~\citet{cai2019}. 
Given that the covering fraction may have a large uncertainty, more quantitative discussion is challenging. 

\subsubsection{Optically thin scenario}\label{sec:thin_rec}
Assuming the gas is in ionization equilibrium, we can estimate the neutral column density averaged over the area of the halo with the equation given by \citet{henwai2013}, $\langle{N_\mathrm{HI}}\rangle$, 
\begin{equation}
\frac{\langle{N_\mathrm{HI}\rangle}}{10^{17.2}\ \mathrm{cm^{-2}}} = \left( \frac{L_\mathrm{halo}}{10^{44}\ \mathrm{erg\ s^{-1}}} \right)\left(\frac{L_\mathrm{\nu_{LL}}}{10^{30}\ \mathrm{erg\ s^{-1}Hz^{-1}}}\right)^{-1}.
\label{eq:avN}
\end{equation}

The median of the $\langle{N_\mathrm{HI}}\rangle/10^{17.2}\ \mathrm{cm}^{-2}$ based on the measured $L_\mathrm{halo}$ and $L_\mathrm{\nu_{LL}}$ is estimated to be 0.42 and 0.36 at $z\sim4$, and $z\sim6$, respectively, and both satisfy the requirement $\langle{N_\mathrm{HI}}\rangle<10^{17.2}\ \mathrm{cm}^{-2}$ as being optically thin.
However, we should note that $\langle N_{\mathrm{HI}} \rangle < 10^{17.2}\ \mathrm{cm^{-2}}$ does not give a strict indication that the entire \Lya halo is optically thin since local optically thick regions may exist~\citep{henwai2013}. 
The \Lya surface brightness for the case of highly-ionized optically thin gas is given by \citet{henwai2013},
\begin{multline}
  \mathrm{SB} = 9.8 \times 10^{-16}\\
  \times (1+z)^{-4}\left(\frac{f_c}{0.5}\right)\left(\frac{n_\mathrm{H}}{1\ \mathrm{cm}^{-3}}\right)\left(\frac{N_\mathrm{H}}{10^{20.5}\ \mathrm{cm}^{-2}}\right) \ \mathrm{erg\ s^{-1}cm^{-2}arcsec^{-2}} \label{eq:thinsb}
\end{multline}
where $n_\mathrm{H}$ and $N_\mathrm{H}$ are the cloud's hydrogen volume and column densities. 
If we assume the same column density as that of $z\sim2\text{--}3$, $N_\mathrm{H}=10^{20.5}\ \mathrm{cm^{-2}}$ within an impact parameter of 200 pkpc \citep[e.g.][]{lau2016}, high gas density, $n_\mathrm{H}>1\ \mathrm{cm^{-3}}$, is required to explain the \Lya SB observed in this study with Equation (\ref{eq:thinsb}).
Such high density is proposed to explain the emission of the giant \Lya halo at $z\sim2\text{--}3$ \citep{cantalupo2014,battaia2015a,battaia2015b,battaia2018,hennawi2015,cai2018}.
Therefore, we conclude that the observed \Lya emission can be explained consistently by recombination radiation from optically thin gas.

\subsection{Scattering from the broad line region}\label{sec:scattering}
\Lya photons produced in the central Broad Line Region (BLR) and resonantly scattered by the neutral gas in the CGM may also contribute to the emission.
According to~\citet{pezzuli2019}, in optically thick gas, the recombination radiation, which has a larger contribution than scattering, becomes dominant.
However, the difference between the two gradually decreases as it approaches the optically thin limit, and in the thin limit, scattering is expected to be $\sim1$ dex brighter than recombination.
As discussed in Section~\ref{sec:recombination}, the observed SB of \Lya haloes is reasonably explained by the optically thin case, so it is worth discussing whether scattering contributes to the emission. 

\citet{henwai2013} derives the SB averaged over the entire halo produced by scattering,
\begin{multline}    \mathrm{SB} = 4.3 \times 10^{-17} (1+z)^{-4}\left(\frac{f_c}{0.5}\right) \left(\frac{R}{100\ \mathrm{kpc}}\right)^{-2} \\
\times \left(\frac{L_\mathrm{{\nu_{Ly\alpha}}}}{10^{31}\ \mathrm{erg\ s^{-1}\ Hz^{-1}}}\right)\ \mathrm{erg\ s^{-1}\ cm^{-2}\ arcsec^{-2}} \label{eq:scatsb}
\end{multline}
where
$L_\mathrm{\nu_{Ly\alpha}}$ is the intrinsic luminosity density in \Lya. 
Assuming that $L\mathrm{_{\nu_{Ly\alpha}}}$ can be represented by $L\mathrm{^{peak}_{Ly\alpha;QSO}}$, from Equations (\ref{eq:sb2l}) and (\ref{eq:scatsb}), we obtain the following equation;
\begin{equation}
    L_\mathrm{halo} = 6.9\times 10^{42} \left(\frac{f_c}{0.5}\right)\left(\frac{L\mathrm{^{peak}_{Ly\alpha;QSO}}}{10^{31}\ \mathrm{erg\ s^{-1}Hz^{-1}}}\right)\ \mathrm{erg\ s^{-1}}.
 \label{eq:scatL}
\end{equation}
In this case, the \Lya halo luminosity is proportional to the quasar \Lya luminosity.
Figure~\ref{fig:Ldep} shows that the \Lya halo luminosity is linearly correlated to the quasar \Lya luminosity, though its correlation slope is shallower than proportional, as has been seen in the correlation between $L_\mathrm{\nu_{LL}}$ and $L_\mathrm{halo}$.
\citetalias{mackenzie2021revealing} also find a shallow linear relation between $L_\mathrm{\nu_{LL}}$ and $L_\mathrm{halo}$.
A deviation from strict proportionality suggests that the \Lya halo luminosity cannot be explained purely by scattering alone, and is likely to be due to the contribution of recombination radiation from both optically thick (Section~\ref{sec:thick_rec}) and thin (Section~\ref{sec:thin_rec}) clouds.
Furthermore, by using our median values $L\mathrm{^{peak}_{Ly\alpha;QSO}}=3.0\times 10^{31}, 7.3\times 10^{31}\ \mathrm{erg\ s^{-1} Hz^{-1}}$ at $z\sim4$ and $z\sim6$, the \Lya halo luminosities can be estimated from Equation (\ref{eq:scatL}) to be 
$L_\mathrm{halo} \ga 2.1\times 10^{43} , 5.0\times 10^{43} \ \mathrm{erg\ s^{-1}}$, when using $f_c \ga 0.5$, following \citet{battaia2015a,battaia2015b,hennawi2015}. 
This is consistent with our median value of $L_\mathrm{halo}=3.9\times 10^{43} \mathrm{\ erg\ s^{-1}}$ at $z\sim4$. At $z\sim6$, the lower limit is about 1.4 times larger than our median value of $L_\mathrm{halo}=3.5\times 10^{43} \mathrm{\ erg\ s^{-1}}$; however, considering that the above estimate is based on several assumptions, the difference is not significant enough to rule out scattering.

The scattering contribution can also be inferred from the shape of the radial profile. Our SB profiles seen in both Figure~\ref{fig:radial4} and Figure~\ref{fig:radial6} 
can be approximately described by exponential curves and appear flatter in the inner regions compared to a power law. 
A hint of flattening at $\la$ 5 kpc can also be seen in Figure~\ref{fig:each_profile} at both $z\sim4$ and $z\sim6$.
According to \citet{costa2022} based on cosmological radiation-hydrodynamic simulations, scattering is necessary to reproduce this flatter shape. Due to the high H{\sc i} column density at the centre of the host galaxy, \Lya photons are resonantly trapped in optically thick gas and repeatedly scattered in spatial and frequency space until they are enough away from the galactic centre and the frequency shift is large enough to reduce the absorption cross-section. The central flattening of the profile may therefore be due to the presence of H{\sc i}-rich regions at the galactic centre. 
Furthermore, the simulations by \citet{costa2022} have shown that, compared to brighter quasars, the lower luminosity quasars have weaker radiative pressure, so that the gas containing a large amount of dust remains around the centre of the galaxy, and the efficient escape of recombination and BLR photons is prevented, resulting in more pronounced flattening.
\citet{costa2022} also suggest that there could be an optically thick gas filament along with an inflow that extends outwards, through which the scattered \Lya photons reach the outside, contributing to the \Lya halo luminosity.
They suggest that to explain \Lya halo with large extent ($\ga 60\ \mathrm{kpc}$), \Lya photons must be transported outwards by scattering. They also note that if scattering contributes, the halo becomes more asymmetric. 
Although it is difficult to accurately explore the morphology because our data do not have resolved spatial information, the asymmetry is certainly observed in the MUSE observations at $z\sim4$~\citep{christ2006,borisova2016ubiquitous,travascio2020} and $z\sim6$~\citep{roche2014,farina2019requiem}.

It can be concluded that scattering contributes to the radiation from \Lya halo, which changes from optically thick to thin from the inside to the outside.
This is consistent with the discussion of recombination radiation observed in both optically thin and thick cases in Section~\ref{sec:recombination} and provides a coherent explanation for our observational results.



\subsection{Collisional excitation}
As the collisionally excited hydrogen returns to its ground state, \Lya photons are emitted.
When the electron temperature is $2\text{--}5\times 10^4\ \mathrm{K}$, the radiation by collisional excitation exceeds the recombination radiation~\citep{cantalupo2008}. However, the collisional excitation coefficient depends exponentially on the temperature and also on the square of the density~\citep[e.g.][]{pezzuli2019,cantalupo2008}, so fine-tuning of gas density and temperature is required for collisional excitation to be effective.
This means that the density and temperature of all \Lya halo must be within a very narrow range. 
It has been suggested that collisional excitation is significantly less than recombination in highly ionized regions such as the galactic centre 
\citep[e.g.][]{cantalupo2008,borisova2016ubiquitous,costa2022}. In \citet{costa2022}, both recombination and collisional excitation are required to reproduce the radiation, but collisional excitation becomes dominant outside $\sim$30 pkpc from the centre (see Figure 5 in \citealt{costa2022}).
Given that we are looking much further inwards than $r = 30$ pkpc (see Figure~\ref{fig:radial4},~\ref{fig:radial6}), we are not able to, unfortunately, constrain the contribution of collisional excitation by this study.

\subsection{A possible interpretation of correlation between $L\mathrm{_{halo}}$ and $M\mathrm{_{BH}}$
}\label{sec:disc_bh}

In Section~\ref{sec:result_bh}, $L\mathrm{_{halo}}$ dependence on $M\mathrm{_{BH}}$ is found only at $z\sim6$.
As we have already discussed, we should first note the uncertainties of this result, especially the small number of heavy $M\mathrm{_{BH}}$ measurements at $z\sim4$, and that the $M\mathrm{_{BH}}$ measurements for faint quasars at $z\sim6$ are not made by a conservative single-epoch method. 
Nevertheless, we discuss below what this result could suggest.
As discussed in Sections~\ref{sec:recombination} and~\ref{sec:scattering}, the \Lya halo emission, $L\mathrm{_{halo}}$ is contributed from recombination radiation from optically thick gas and thin gas, and scattering both at $z\sim4$ and $z\sim6$.
In the case of recombination radiation from optically thick gas,  
$L\mathrm{_{halo}}$ depends on 
$L_\mathrm{{\nu_{LL}}}$ 
(see Section~\ref{sec:thick_rec}).
We derive $L_\mathrm{{\nu_{LL}}}$ from the UV luminosity (Equation~\ref{eq:M14502M912}), which is correlated with the $M\mathrm{_{BH}}$ based on the single-epoch method~\citep{vestergaard2006}. 
Therefore, the positive correlation between $L\mathrm{_{halo}}$ and $M_\mathrm{BH}$ is naturally expected, but it is difficult to explain why no correlation is seen at $z\sim4$.
Similarly in the case of scattering, it is also possible that the $L\mathrm{_{halo}}$-$L^{\mathrm{peak}}_{\mathrm{Ly\alpha;QSO}}$ relation and the loose correlation between $L^{\mathrm{peak}}_{\mathrm{Ly\alpha;QSO}}$ and UV luminosity (\citetalias{mackenzie2021revealing}) produces the $L\mathrm{_{halo}}$-$M\mathrm{_{BH}}$ relation, but the result that the correlation is only seen in $z\sim6$ calls into question its plausibility.
In the case of recombination radiation from optically thin gas, $L\mathrm{_{halo}}$ depends on the $n_{\mathrm{H}}$ and $N_{\mathrm{H}}$ (see Section~\ref{sec:thin_rec}). 
The observed positive correlation between $L\mathrm{_{halo}}$ and $M_\mathrm{BH}$ at $z\sim6$ can be expected, assuming that galaxies that have taken longer to foster its heavy central SMBH also have higher $N_{\mathrm{H}}$, due to the cumulative amount of gas accretion from the surrounding IGM.
The relationship is maintained in the early universe at $z\sim6$, but 600 Myr later at $z\sim4$, it may have faded due to the complex baryon physics processes in the CGM cold gas, such as repeatedly consumption for star formation, and blowing out by outflows.
Although it is difficult to make a direct comparison with our results, ~\citet{momose2019} claims that there is a moderate correlation between $M\mathrm{_{BH}}$ and $L\mathrm{_{halo}}/L_{\mathrm{bol}}$ at $z\sim2\text{--}3$.

The $\lambda\mathrm{_{edd}}$ is proportional to $L_{\mathrm{bol}}/M_{\mathrm{BH}}$, where $L_{\mathrm{bol}}$ is the bolometric luminosity of the quasar and it is related to the UV luminosity. 
It is reasonable that there is no correlation between $\lambda\mathrm{_{edd}}$ and $L\mathrm{_{halo}}$ at $z\sim6$, where $L\mathrm{_{halo}}$ is correlated with both $M_\mathrm{BH}$ and the UV luminosity.
At $z\sim4$, where $L\mathrm{_{halo}}$ is not correlated with $M_\mathrm{BH}$ but only with the UV luminosity, an inverse correlation between $\lambda\mathrm{_{edd}}$ and $L\mathrm{_{halo}}$ is expected, but such a correlation is not found.
In any case, there is an underlying relation between the observables being discussed here, so it is difficult to find the essential correlation.

\section{SUMMARY AND CONCLUSIONS}
We search the \Lya haloes around faint quasars at $z\sim4$ and $z\sim6$ based on long-slit spectroscopy datasets. 
Our parent sample consists of 20 quasars at $3.45 \leq z \leq 4.10$ taken by \citet{he2024black} and 162 $5.66 \leq z \leq 7.07$ quasars taken by SHELLQs, and after careful PSF subtraction, 12 and 26 \Lya haloes are detected at $z\sim4$ and $z\sim6$, respectively. 
The detected sample has an average absolute magnitude of $\langle M_{1450} \rangle = -23.84$ mag at $z\sim4$, which is $\sim 4$ mag fainter than the previous study by \citetalias{borisova2016ubiquitous} and comparable to \citetalias{mackenzie2021revealing}, and $\langle M_{1450} \rangle = -23.68$
mag at $z\sim6$, which is $\sim3$ mag fainter than \citetalias{farina2019requiem}. Our main findings are as follows:
\begin{enumerate}
    \item
    Both at $z\sim4$ and $z\sim6$, the median SB profiles are consistent with the exponential curve, $(1+z)^4\mathrm{SB} = C \exp{(-r/r_h)}$, though our sample only traces the inner region at $\la$ 10 pkpc of \Lya halo. 
    The SB profiles show the flattening at $\la$ 5 pkpc.
    The \Lya haloes around these faint quasars are found to be systematically fainter than those around bright quasars in the previous studies.
    \item
    The total luminosity of \Lya haloes is positively correlated with the quasar luminosities both at the Lyman limit and at the peak of the \Lya line, the latter being stronger. Such a luminosity dependence at $z\sim6$ is first found in this study by expanding the dynamic range of quasar luminosities.
    \item The correlation between \Lya halo luminosity and quasar ionizing luminosity suggests a contribution of recombination radiation from optically thick gas. However, fully optically thick gas alone exceeds the observed \Lya halo emission, implying the presence of optically thin regions.
    \item 
    The observed \Lya halo luminosity can be explained consistently by recombination radiation from optically thin gas when assuming the same column density of $N_\mathrm{H}=10^{20.5}\ \mathrm{cm^{-2}}$ and the high gas density of $n_{\mathrm{H}}>1\ \mathrm{cm^{-3}}$ as those at $z\sim2\text{--}3$.
    \item 
    The positive correlation between the \Lya halo luminosity and quasar peak \Lya luminosity implies a contribution of resonant scattering. The shape of the radial profile, which flattens at the centre, also indicates that scattering contributes to the radiation from \Lya halo, which changes from optically thick to thin from the inside to the outside. 
    \item 
    The \Lya halo luminosity is positively correlated with SMBH mass at $z\sim6$, while not at $z\sim4$.
    No statistically robust correlation is found between \Lya halo luminosity and SMBH Eddington ratio, either at $z\sim4$ or at $z\sim6$.
\end{enumerate}

The observed SB of the halo can be reasonably explained by optically thin recombination, but if we also consider the two additional contributions of optically thick recombination to explain the dependence of \Lya halo luminosity on $L_\mathrm{\nu_{LL}}$  and scattering to explain its dependence on $L\mathrm{^{peak}_{Ly\alpha;QSO}}$, we can explain everything in a consistent manner.
Although the CGM structure cannot be determined from this observation, when compared with the hydro-dynamical simulation \citep{costa2022}, the CGM transitions from the optically thick inner region to the optically thin outer region, and the recombination radiation from each location is reasonably possible. 
In addition, if we consider that the optically thick filamentary structure of the inflow is also present, the scattering radiation can be fully explained. 
This implies a picture of the complex structure and radiation transfer in the CGM halo.
If IFU observations with higher spatial resolution can be made in the future, it may be possible to resolve these complications by examining, for example, the luminosity dependence for each part of the halo.
Our finding of \Lya halo luminosity dependence on the quasar luminosity at both $z\sim4$ and $z\sim6$ places constraints on the powering mechanisms of the \Lya halo.
Interestingly, a similar dependence on quasar ionizing luminosity is found in quasars at $z=2\text{--}3$ by \citet{shimakawa2022}, who finds a higher detection rate of \Lya halo in more luminous quasars.
It is necessary to investigate more systematically whether this luminosity dependence exists by increasing the number of faint quasars over cosmic time.
Since this study used existing slit-spectroscopic data, it is difficult to trace individual radial profiles far enough outside of the halo.
Nevertheless, this study demonstrates that even long-slit data can provide some insights into the \Lya halo.
We expect to be able to obtain more detailed SB profiles and improve the accuracy of the luminosity dependence correlations by increasing the number of samples with long exposures in the future.
Further studies of the spatial extent of other emission lines, such as H$\alpha$, which is now feasible with JWST, can further constrain the powering mechanisms.
Comparing the radial profiles of H$\alpha$ and \Lya can help constrain to what extent scattering contributes because only \Lya photons undergo resonant scattering~\citep{masribas2017}. 
If the emission mechanisms are fully elucidated, we will be able to gain detailed insights into the physical properties of the CGM through \Lya haloes, representing significant progress in CGM research.

\section*{Acknowledgements}
We appreciate the anonymous reviewer for constructive
comments and suggestions.
We thank Tiago Costa for sharing their results and providing useful suggestions for this study.
We appreciate Kazuhiro Shimasaku for constructive discussions. NK was supported by the Japan Society for the Promotion of Science through Grant-in-Aid for Scientific Research 21H04490. YM was supported by the JSPS KAKENHI grant No. 21H04494.  SK was supported by the JSPS KAKENHI grant No. 24KJ0058, 24K17101. KI acknowledges support under the grant PID2022-136827NB-C44 provided by MCIN/AEI/10.13039/501100011033 / FEDER, UE.


\section*{Data Availability}
The data in this paper will be shared upon reasonable request.
 



\bibliographystyle{mnras}
\bibliography{reference}




\appendix

\section{One-dimensional spectra}\label{full_oned}
\begin{figure*}  
    \centering
    \foreach \img in {z4_1_oned.png,z4_0_oned.png,z4_5_oned.png,z4_4_oned.png,z4_3_oned.png,z4_6_oned.png,z4_12_oned.png,z4_9_oned.png,z4_8_oned.png,z4_10_oned.png,z4_14_oned.png,z4_17_oned.png} {%
        \subfloat{\includegraphics[width=0.19\textwidth]{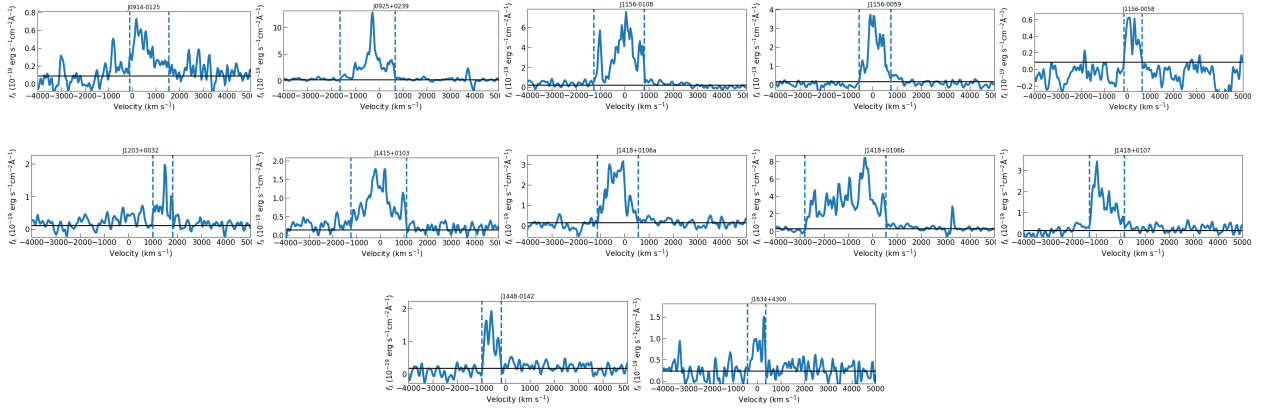}}
        \hspace{-2mm}
    }
    \caption{The one-dimensional spectra after PSF subtraction for the quasars at $z\sim4$. The region between two blue dashed lines corresponds to the wavelengths where the SNR calculated after the first PSF subtraction is larger than two to integrate the residual fluxes. A black horizontal line shows the 1$\sigma$ SB limit.}
    \label{fig:z4detect}
\end{figure*}

\begin{figure*}  
    \centering
    \foreach \img in 
    {z6_29_oned.png,z6_109_oned.png,z6_49_oned.png,z6_62_oned.png,z6_63_oned.png,z6_89_oned.png,z6_136_oned.png,z6_137_oned.png,z6_30_oned.png,z6_133_oned.png,z6_74_oned.png,z6_27_oned.png,z6_112_oned.png,z6_6_oned.png,z6_128_oned.png,z6_105_oned.png,z6_81_oned.png,z6_72_oned.png,z6_56_oned.png,z6_42_oned.png,z6_20_oned.png,z6_111_oned.png,z6_57_oned.png,z6_0_oned.png,z6_37_oned.png,z6_21_oned.png
}
    {%
        \subfloat{\includegraphics[width=0.19\textwidth]{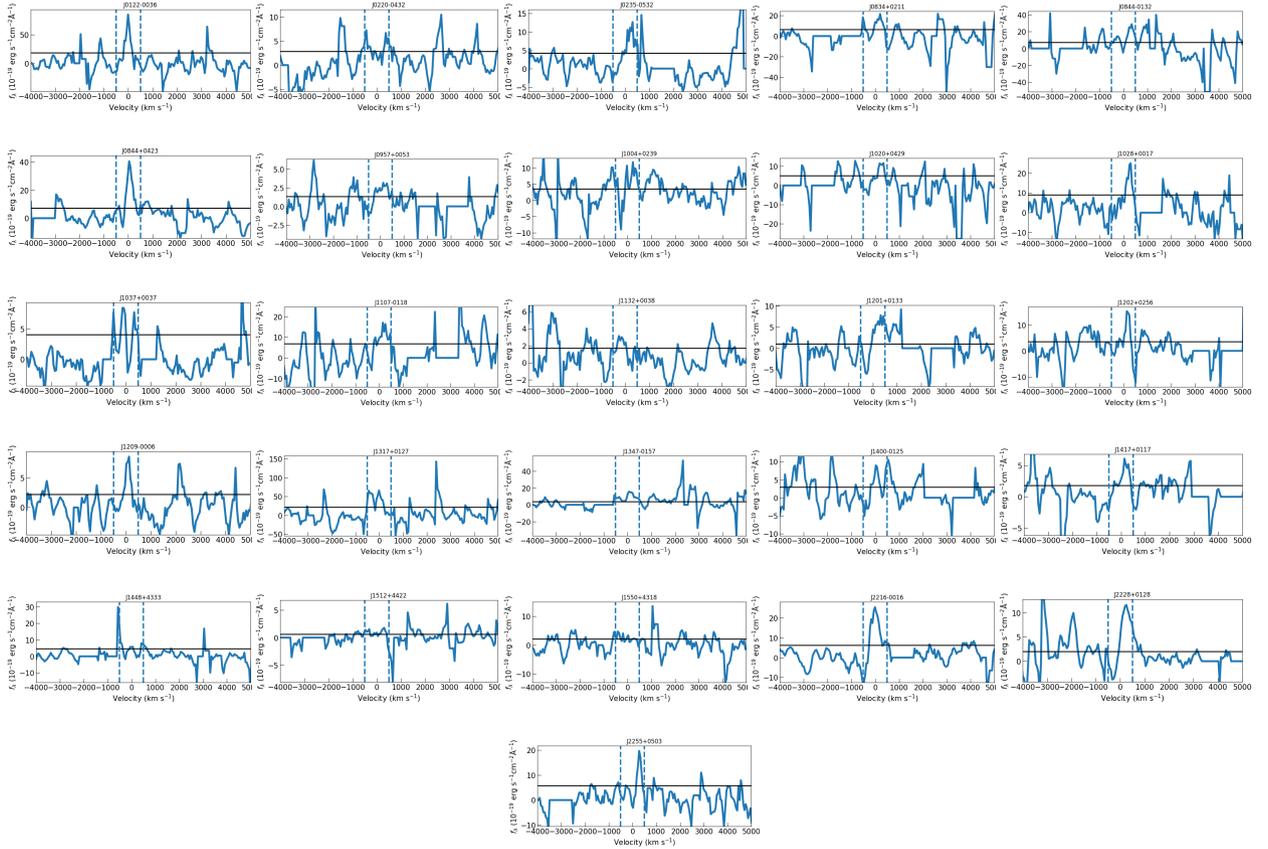}}
        \hspace{-2mm}
    }
    \caption{The one-dimensional spectra after PSF subtraction for the quasars at $z\sim6$ with \Lya halo detection. The region between two blue dashed lines corresponds to the wavelength with $-500<v<500\ \mathrm{km\ s^{-1}}$.}
    \label{fig:z6detect}
\end{figure*}

One-dimensional spectra of PSF-subtracted images are shown with 1$\sigma$ SB limits. 
The SB variance is measured within the same detection spatial and wavelength window.
Figure~\ref{fig:z4detect} shows 12 halo-detected objects at $z\sim4$, Figure~\ref{fig:z6detect} shows 26 objects with \Lya halo detection at $z\sim6$ and Figure~\ref{fig:z6nodetect} shows 86 objects without \Lya halo detection at $z\sim6$. 

\graphicspath{{ap_figures/z4_oned_rev/}}

\graphicspath{{ap_figures/oned_d_v/}}

\graphicspath{{ap_figures/oned_nd_v/}}
\begin{figure*}  
    \centering
    \foreach \img in {z6_148_oned.png,z6_113_oned.png,z6_22_oned.png,z6_35_oned.png,z6_94_oned.png,z6_95_oned.png,z6_96_oned.png,z6_71_oned.png,z6_23_oned.png,z6_31_oned.png,z6_8_oned.png,z6_88_oned.png,z6_10_oned.png,z6_51_oned.png,z6_141_oned.png,z6_52_oned.png,z6_150_oned.png,z6_32_oned.png,z6_33_oned.png,z6_68_oned.png,z6_46_oned.png,z6_43_oned.png,z6_12_oned.png,z6_5_oned.png,z6_44_oned.png,z6_13_oned.png,z6_78_oned.png,z6_102_oned.png,z6_14_oned.png,z6_15_oned.png,z6_53_oned.png,z6_69_oned.png,z6_24_oned.png,z6_115_oned.png,z6_25_oned.png,z6_125_oned.png,z6_127_oned.png,z6_75_oned.png,z6_86_oned.png,z6_26_oned.png,z6_79_oned.png,z6_82_oned.png,z6_121_oned.png,z6_146_oned.png,z6_65_oned.png
}
    {%
        \subfloat{\includegraphics[width=0.19\textwidth]{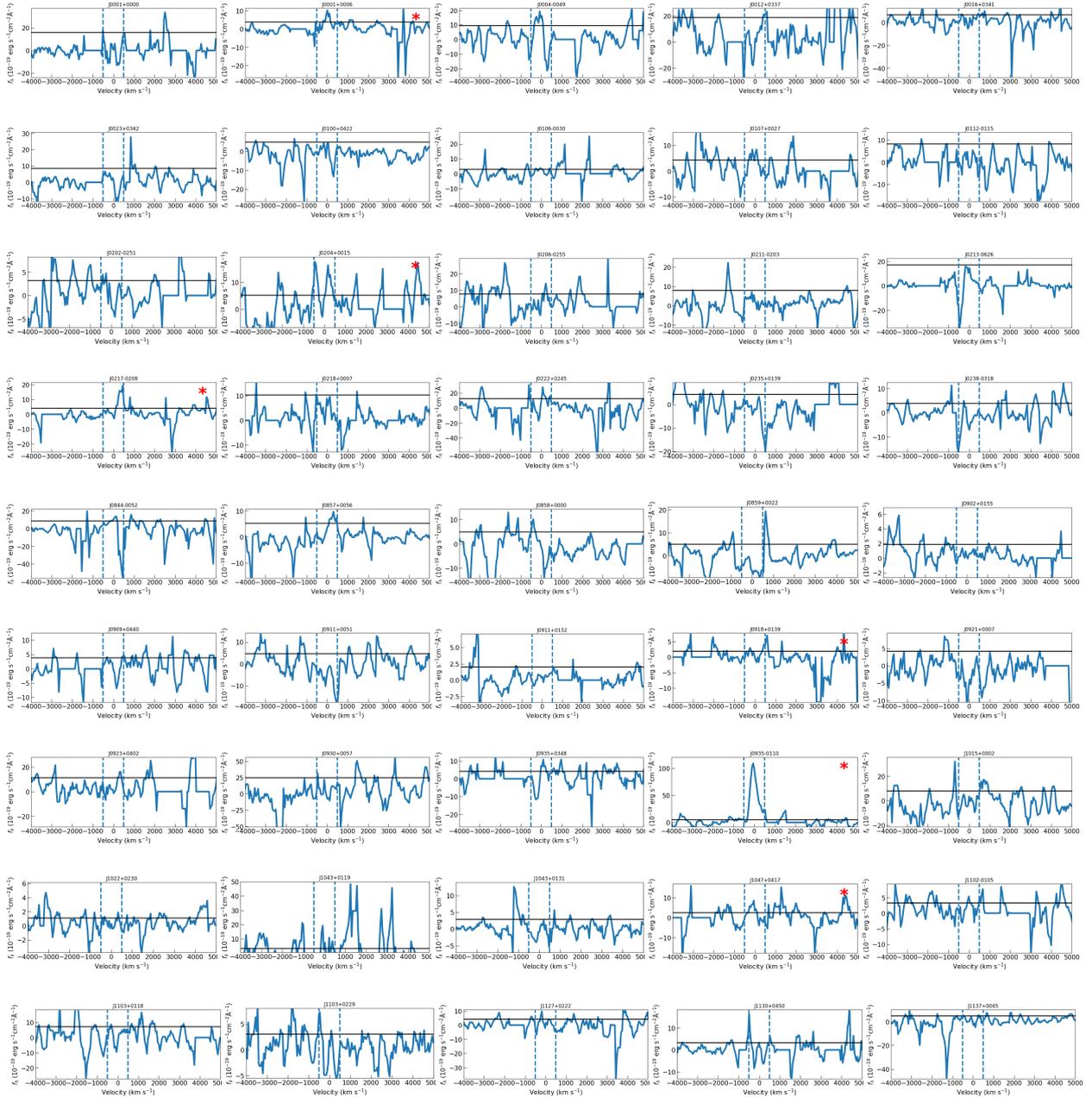}}
        \hspace{-2mm}
    }
    \caption{The one-dimensional spectra after PSF subtraction for the quasars at $z\sim6$ without \Lya halo detection. The red star in the upper right corner indicates that the object is removed by visual inspection.}
    \label{fig:z6nodetect}
\end{figure*}

\begin{figure*}\ContinuedFloat  
    \centering
    \foreach \img in  {z6_54_oned.png,z6_104_oned.png,z6_91_oned.png,z6_2_oned.png,z6_40_oned.png,z6_7_oned.png,z6_110_oned.png,z6_64_oned.png,z6_155_oned.png,z6_80_oned.png,z6_77_oned.png,z6_124_oned.png,z6_28_oned.png,z6_156_oned.png,z6_66_oned.png,z6_55_oned.png,z6_16_oned.png,z6_9_oned.png,z6_98_oned.png,z6_135_oned.png,z6_106_oned.png,z6_70_oned.png,z6_3_oned.png,z6_59_oned.png,z6_4_oned.png,z6_92_oned.png,z6_93_oned.png,z6_19_oned.png,z6_17_oned.png,z6_140_oned.png,z6_138_oned.png,z6_18_oned.png,z6_47_oned.png,z6_39_oned.png,z6_11_oned.png,z6_107_oned.png,z6_61_oned.png,z6_73_oned.png,z6_34_oned.png,z6_117_oned.png,z6_126_oned.png
}
    {%
        \subfloat{\includegraphics[width=0.19\textwidth]{\img}}
        \hspace{-2mm}
    }
    \caption{(Continued.)}
\end{figure*}



\bsp	
\label{lastpage}
\end{document}